\documentclass[10pt,conference,compsocconf,letter]{IEEEtran}

\usepackage{cite}
\usepackage[pdftex]{graphicx}
\usepackage{amsfonts,amssymb,amsmath,alltt}
\usepackage{array}
\usepackage{mdwmath}
\usepackage{mdwtab}
\usepackage{fixltx2e}
\usepackage{stfloats}
\usepackage{url}
\usepackage{xspace}
\usepackage{color}
\usepackage{stmaryrd}

\usepackage{alltt}

\newenvironment{ttbox}{\begin{alltt}\ttbraces\small\tt}%
                      {\end{alltt}}
\def\ttbraces{\let\.=\nobreak\chardef\{=`\{\chardef\}=`\}\chardef\|=`\\}
\newcommand\ttequiv{\mbox{{$\equiv$}}}
\newcommand\ttexists{\mbox{{$\exists$}}}
\newcommand\ttforall{\mbox{{$\forall$}}}
\newcommand\ttneg{\mbox{{$\neg$}}}

\newcommand\ttin{\mbox{{$\in$}}}
\newcommand\ttnin{\mbox{{$\notin$}}}
\newcommand\ttimp{\mbox{{$\longrightarrow$}}}
\newcommand\ttImp{\mbox{{$\Longrightarrow$}}}
\newcommand\ttlam{\mbox{\( \lambda \)}}
\newcommand\ttand{\mbox{{$\land$}}}

\newcommand\ttfun{\mbox{\(\Rightarrow\)}}

\newcommand\tttimes{\mbox{\( \times \)}}

\newcommand\ttrelI{\mbox{{$\to_{i}$}}}
\newcommand\ttrelIstar{\mbox{{$\to_{i}^*$}}}
\newcommand\ttatI{\mbox{\( @_G \)}}

\newcommand\ttvdash{\mbox{{$\vdash$}}}

\newcommand\ttsigma{\mbox{{$\sigma$}}}

\newcommand\ttsubseteq{\mbox{{$\subseteq$}}}

\newcommand\ttref{\mbox{{$\sqsubseteq$}}}

\newcommand{\ttcalN}[1]{\mbox{{${\cal{N}}_{\texttt{#1}}$}}} 
\newcommand\ttattand[1]{\mbox{{$\oplus_{\wedge}^{#1}$}}}
\newcommand\ttattor[1]{\mbox{{$\oplus_{\vee}^{#1}$}}}
\newcommand\ttrel[1]{\mbox{{$\to_{#1}$}}}
\newcommand\ttrelstar[1]{\mbox{{$\to_{#1}^*$}}}

\newcommand\tcon{\ \hat{}\ }
\newcommand\choi{\, \Box \,}
\newcommand\ttmref[1]{\mbox{{$\sqsubseteq_{#1}$}}}
\newcommand\ttmeref{\ttmref{\mathcal{E}}}
\newcommand\ttecal{\mbox{$\mathcal{E}$}}
\newcommand\ttimg{\mbox{\texttt{`}}}
\newcommand\ttbigcap{\mbox{{$\bigcap$}}}
\newcommand\ttgeq{\mbox{$\geq$}}
\begin{document}
\title{Dependability Engineering in Isabelle}
\author{\IEEEauthorblockN{Florian Kamm\"uller}
\IEEEauthorblockA{Department of Computer Science\\Middlesex University London\\
f.kammueller@mdx.ac.uk}
}
\maketitle              % typeset the title of the contribution

\begin{abstract}
In this paper, we introduce a process of formal system development supported by
interactive theorem proving in a dedicated Isabelle framework. This Isabelle Infrastructure
framework implements specification and verification in a cyclic process supported by attack
tree analysis closely inter-connected with formal refinement of the specification. The process
is cyclic: in a repeated iteration the refinement adds more detail to the system specification.
It is a known hard problem how to find the next refinement step: this problem is addressed by
the attack based analysis using Kripke structures and CTL logic. We call this cyclic process the
Refinement-Risk cycle (RR-cycle). It has been developed for security and privacy of IoT healthcare
systems initially but is more generally applicable for safety as well, that is, dependability
in general.
In this paper, we present the extensions to the Isabelle Infrastructure framework implementing a
formal notion of property preserving refinement interleaved with attack tree analysis for the
RR-cycle. The process is illustrated on the specification development and privacy analysis of the
mobile Corona-virus warning app.
\end{abstract}

\section{Introduction}
\label{sec:intro}

\subsection{Motivation}
Dependability \cite{alrl:04} is a notoriously difficult property for system development because already
security -- which is part of dependability -- is not compositional: 
given secure components, a system created from those components is not necessarily secure.
Therefore, the usual divide-and-conquer approach from system and software design does not
apply for dependability engineering. At the same time, it is mandatory for the design of
dependable systems to introduce security in the early phases of the development since it
cannot be easily “plugged in” at later stages. However, even if security is introduced in
early phases, a classical stepwise development of refining abstract specifications by concretizing
the design does not preserve security properties. Take for example, the implementation of sending
a message from a client A to a server B such that the communication is encrypted to protect its
content. In the abstract system specification, we do not consider a concrete protocol nor the
architecture of the client and server. Using common refinement methods from software engineering
provides a possible implementation by passing the message from a client system AS connected by a
secure channel to a server system BS. However, this implementation does not exclude that other
processes running on either AS or BS can eavesdrop on the plain text message because the confidentiality
protection is only on the secure channel from AS to BS. This example is a simple illustration of
what is known as the security refinement paradox \cite{jac:89}.

Why is security so hard? A simple explanation is that it talks about negative properties: something
(loss or damage of information or functionality) must not happen. Negation is also in logic a difficult
problem as it needs exclusion of possibilities. If the space to consider is large, this proof can be hard
or infeasible. In security, the attacks often come from ``outside the model''. That is, for a given specification
we can prove some security property and yet an attack occurs which uses a fact or observation or loophole
that just has not been considered in the model. This known practical attack problem is similar to the
refinement paradox. Intuitively, the attacker exploits a refinement of the system that has not been
taken into account in the specification but is actually part of the real system (an implementation of
the specification). In the above example, the real system allows that other applications can be run on
the client within the security boundary. This additional feature of multi-processing systems has not
been taken into account in the abstract specification where we considered processes and systems - the
client and the server - as abstract entities without considering crucial features of their internal
architectures.

To facilitate the formal stepwise development of dependable systems, we propose to make use of these
intricacies by including the attacks into the process. As in common engineering practice, where testing
for failures is used for improvement of prototypes, we use attack tree analysis within
the same formal framework to provide guidance for the refinement of the system specification.
Since the interleaved attack analysis steps are formalized in the same expressive formal framework,
the found attack paths lead to system states that provide concrete information about the loopholes.
This information allows adding details to data types, system functions, as well as policies thereby
providing a next level of a more refined specification closer to an implementation. This is a cyclic
process: in turn the newly refined specification can be scrutinized by attack tree analysis since
the added detail may provide additional new loopholes. However, there is a stepwise improvement.
Since we are using the highly expressive Higher Order Logic of Isabelle as a basis for our formal
framework, we can implement the framework so that it allows expressing infrastructures including actors
and policies, attack trees based on Kripke structures and temporal logic CTL, as well as a closed
notion of dependability refinement that encompasses the idea that a refined specification after attack
tree analysis excludes previous attacks.

%However, these intricacies of security refinement and preservation of security do not occur so much,
%if we lift the negation. Instead of considering the negative security property, we double negate and
%consider ``not secure''. In other words, we contemplate attack possibilities.
%The working hypothesis of this paper is that we can have stronger refinement properties when
%considering attacks. What we want to examine is whether attacks can be refined for given system models.
%We will do this by first providing a formal model of attacks using the well known attack tree concept
%but under-laying it with a formal foundation by providing a logical proof theory for it. Refinement is
%in fact the natural way of proving an attack on a given system by applying the calculus to find a
%concrete way that realizes an abstract attack. The attack refinement shows how the concrete
%implementation must look like in order to enable the attack. Thus, ultimately, we find a negative –
%not secure implementation of an abstract system – by letting the attack lead the way of the refinement.

\subsection{Contributions}
This paper %presents for the first time
for the first time completes the RR-cycle by presenting
(a) a full formal definition of the RR-cycle including
(b) a formal notion of refinement with a property preservation theory, and
(c) a formal notion of termination of the RR-cycle.
All of those definitions and related theorems are validated on the application example of the Corona-cirus Warning App (CWA) 
completing a preliminary workshop publication \cite{kl:20}.
It thus completes and extends previous works on establishing the RR-cycle. A more detailed account
of this completion and how it relates to previous works is given in Section \ref{sec:relisa}.

\subsection{Overview of the paper}
In this paper, we first give a general overview over different approaches to refinement
in formal methods in Section \ref{sec:refcomp} to provide a better understanding and foundation
for our formal framework. In Section \ref{sec:ref}, we then proceed to present the technical core
of the paper, the formal definition of refinement including property preservation and mainly the
definition of the RR-Cycle itself by providing a termination condition.
Section \ref{sec:cwa} introduces the case study, the CWA.
Then, we illustrate application of the Isabelle Infrastructure framework for modeling and initially
analyzing the application in Section \ref{sec:model}.
The methodology of the RR-Cycle is illustrated on the CWA case study in Section \ref{sec:cwaref}. That
is, we repeatedly refine the initial specification based on the attack tree results until the termination
condition of the RR-Cycle is met and the global policy is thus valid.
Section \ref{sec:related} discusses related work and presents conclusions.
An overview over the Isabelle Infrastructure framework that is used as the vehicle 
for the formalisation of the RR-cycle is given next in Section \ref{sec:infra}.
%\subsubsection{Contents}

\section{Isabelle Infrastructure Framework}
\label{sec:infra}
\begin{figure}
\vspace{-.5cm}
  \begin{center}
  \includegraphics[scale=.45]{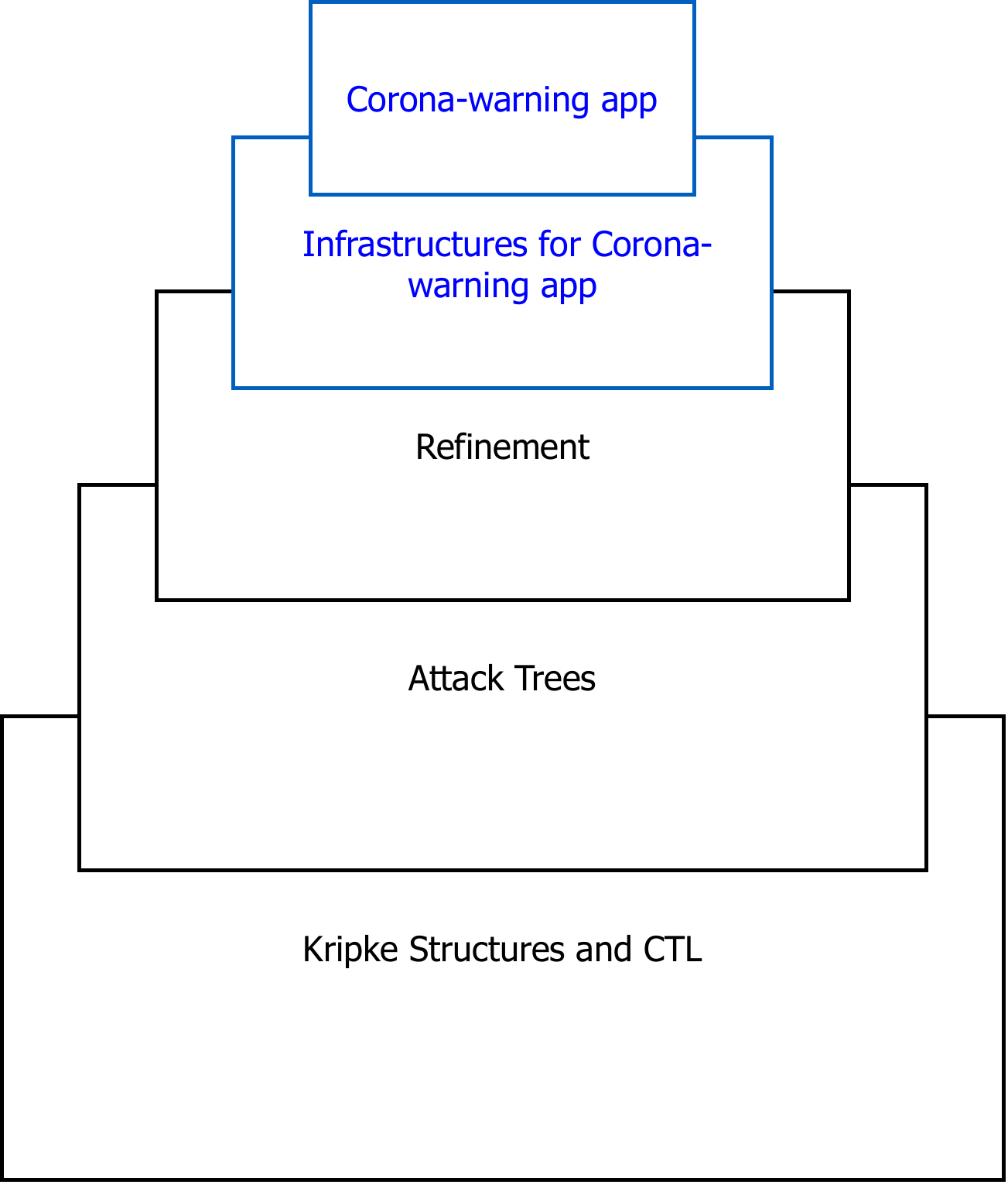}
\end{center}
\vspace{-.5cm}
\caption{Structure of Isabelle Infrastructure framework for application CWA.}
\label{fig:theorystruc}
\vspace{-.5cm}
\end{figure}

%\TODO{Adapt: this is copied from FMBC paper}
The Isabelle Infrastructure framework is implemented as an instance of
Higher Order Logic in the interactive generic theorem prover Isabelle/HOL \cite{npw:02}.
The framework enables formalizing and proving of systems with physical and logical components,
actors and policies. It has been designed for the analysis of insider 
threats. However, the implemented theory of temporal logic combined with Kripke structures and its
generic notion of state transitions are a perfect match to be combined with  attack trees into
a process for formal security engineering \cite{suc:16} including an accompanying framework \cite{kam:19a}.
%\subsubsection{Overview}
%\label{sec:isainfover}
An overview of the layers of object-logics of the Isabelle Infrastructure framework 
is provided in Figure \ref{fig:theorystruc}: each level above is embedded into the one below; the novel
contribution of the Corona-virus warning app that will be presented in this paper are the highlighted
two top boxes 
%in blue on the high end
of the figure.

\subsubsection{Kripke structures, CTL and Attack Trees}
%\label{sec:intro}
A number of case studies (see Section \ref{sec:related})
have contributed to shape the Isabelle framework into a general framework for
the state-based security analysis of infrastructures with policies and actors. Temporal logic
and Kripke structures are deeply embedded into Isabelle's Higher Order logic thereby
enabling meta-theoretical proofs about the foundations: for example, equivalence between attack trees 
and CTL statements have been established \cite{kam:18b} providing sound foundations for applications.
%As part of the meta-theory Correctness and Completeness have been 
%proved in Isabelle \cite{kam:18b} and can be used to navigate between CTL and attack 
%trees to find attacks. 
This foundation provides a generic notion of state transition on which attack trees and
temporal logic can be used to express properties for applications. 
The logical concepts and related notions thus provided for sound application modeling are:
\begin{itemize}
\item {Kripke structures and state transitions:}\\ 
A generic state transition relation is $\ttrelI$; Kripke structures
over a set of states \texttt{t} reachable by $\ttrelI$ from an initial 
state set \texttt{I} can be constructed by the \texttt{Kripke} contructor as
\begin{ttbox}
Kripke \{t. \ttexists i \ttin I. i \ttrelIstar t\} I
\end{ttbox}
\item {CTL statements:}\\ 
We can use the Computation Tree Logic (CTL) to specify dependability properties as 
\begin{ttbox}
K \ttvdash {\sf EF} s
\end{ttbox}
This formula states that in Kripke structure \texttt{K} there is a path ({\sf E}) on which
the property \texttt{s} (given as the set of states in which the property is true) will eventually ({\sf F}) hold.
\item {Attack trees:} \\
attack trees are defined as a recursive datatype in Isabelle having three constructors: 
$\oplus_\vee$ creates or-trees and $\oplus_\wedge$ creates 
and-trees.
And-attack trees $l \ttattand s$ and or-attack trees $l \ttattor s$ 
consist of a list of sub-attacks which are themselves recursively given as attack trees.
The third constructor takes as input a pair of state sets constructing a base attack step between
two state sets. For example, for the sets \texttt{I} and \texttt{s} this is written as \texttt{\ttcalN{(I,s)}}.
As a further example, a two step and-attack leading from state set \texttt{I} via \texttt{si} to \texttt{s} is
expressed as 
\begin{ttbox}
\ttvdash [\ttcalN{(I,si)},\ttcalN{(si,s)}]\ttattand{\texttt{(I,s)}}
\end{ttbox}
%\begin{ttbox}
%\ttvdash [\ttcalN{(I,s)},\ttcalN{(s,s')}]\ttattand{(I,s')}
%\end{ttbox}
\item {Attack tree refinement, validity and adequacy:}\\
Attack trees can be constructed also by a refinement process but this is different from
the system refinement presented in this paper (see Section \ref{sec:ref}).
An abstract attack tree may be refined by spelling out the attack steps until a valid attack
is reached:

\texttt{\ttvdash A :: (\ttsigma :: state) attree)}.

The validity is defined constructively so that code can be generated from it.
Adequacy with respect to a formal semantics in CTL is proved and can be used to
facilitate actual application verification. This will be used for the stepwise system refinements
central to the methodology presented in this paper.
\end{itemize}

\section{Formal Refinement Overview}
\label{sec:refcomp}
%An additional feature that has been integrated into the Isabelle Infrastructure
%framework motivated by security engineering formal specifications for IoT healthcare
%system is an extension of the formal specification process introducing refinement of
%Kripke structures \cite{kam:19a,kam:20a}. It refines a system model based on a formal definition of a
%combination of trace refinement and structural refinement (or datatype refinement).
%The definition allows to prove property preservation results crucial for an iterative
%development process. The refine- ments of the system specification can be interleaved
%with attack analysis while security properties can be proved in Isabelle. In each iteration
%security qualities are accumulated while continuously attack trees scrutinize the design.
%One of the contributions of this paper is to explore different concepts of refinement:
%the formal expression of refinement, enables to pin down (i.e. exemplify) different
%concepts of refinement (data refinement, action refinement, trace refinement (aka spec
%refinement) and combinations thereof with concrete attack scenarios.

%Notes: 
Refinement is intuitively thought of as a systematic way of developing system designs from some
abstract specification to a more detailed one. In the process of doing so, details that are left out
in the abstraction are filled in.
%New:
Refinement is a very natural way of step by step engineering systems starting from an abstract system design.
However, whereas in traditional engineering disciplines
%the engineering uses rules of thumb and is validated through
validation is done through simulation, prototyping and testing, software-based systems (including IoT systems)
allow formally specifying abstract system designs with logic. Then the concept of refinement can be explicitly
formalized providing means to prove properties of the formal system design (also known as system specification)
as well as reasoning in general about refinement, most importantly, how the refinement preserves properties
or achieves design goals.

% From notes (on systems and specification): could go to intro of other section
We consider systems as IoT system or cyber-physical systems, that is, systems constituted of hardware and
software components. We emphasize that these systems should be designed in a human centric way. Although
humans cannot be considered as system parts, their interaction can be specified as well by considering
policies that specify the possible behaviour of actors as parts of the system design. If behavioural descriptions
of human behaviour exist, they may well be integrated into the system specification.
An example are insider threats where a taxonomy and empirical knowledge about human behaviour has been
specified to provide a more detailed model of psychological dispositions of actors on when they start being
insider attackers in order to analyze attacks and countermeasures.
System specifications consist of datatypes for actors, physical system aspects (hardware, credentials, locations)
and logical aspects, e.g., data, access control labels, software functions (types of functions, other descriptions
of functionality like logical (axiomatic) descriptions) and policies.
To summarize this heterogeneous set of entities, we use the simple description ``infrastructures with (human)
actors and policies''.

% Bit of a repetition?
Refinement is the process that starts with an initial more or less formal description of such a system
adding more detail. This process is iterative, best described as a cycle in which a range of design and
analysis activities lead to a more definite, concrete specification and thus a more specific, tangible
description of the system.

\subsection{Refinement methods}
The concept of formal refinement has a long standing in software system engineering in
communicating and reactive system methods, for example, state charts, CSP \cite{hoa:85},
CCS \cite{mil:80a}, the Pi-calculus \cite{mil:99}, as well as in data-based system specification methods
like Z \cite{abr:74,spy:89}, B \cite{abr:96}, VDM. The field is well-explored and the literature is consequently vast so it
is beyond the scope of this paper to give a comprehensive survey.
Nevertheless, we are providing here an informal overview of a range of major techniques commonly used in
these formal methods. This overview serves conveying the conceptual ideas to prepare the way for their
implementation in the formal refinement of the RR-cycle and illustrate them on the application. The related
work, Section \ref{sec:related}, will, however, contrast in more detail to similar formal refinement approaches.
%Before introducing the formal definition of system refinement, we provide an informal motivation of
%what refinement is and what different kinds of refinement exist.

\subsubsection{State based trace refinement of Isabelle Infrastructure framework}
The Isabelle Infrastructure framework enables a state-based specification (sometimes called model-driven)
of systems where the overall system behaviour is defined (by an inductive definition) as a set of traces of system
states (that include all aspects of infrastructures, actors, and policies). The system behaviour is the set
of all possible traces. The general idea of refinement is to make the behaviour more concrete, that is, to
reduce the set of all possible traces. Then a refinement can be simply defined by trace inclusion: in other
words, system specification $S_0$ refines to system specification $S_1$, formally,
\[ S_0 \sqsubseteq S_1 \equiv \texttt{traces}(S_0) \supseteq \texttt{traces}(S_1)\]
(the inversion of the inclusion signs is deliberate not accidental).
It turns out that this inclusion based refinement is quite versatile compared to existing forms of refinement
simply because it combines a rich state based view with the idea of traces of system changes that include
events (or actions) while the events themselves are abstract but can be refined by adding detail to the context
conditions in the inductive definitions associated to the events.
We now consider the various forms of refinement informally to prepare the ground and to give reference
points to later stages when we explain in the application of the Isabelle Infrastructure framework
which form of refinement we are looking at in a specific instance.

\subsubsection{Event based trace refinement}
Formally, the definition of trace refinement is given above: more refined behavioural models have a
subset of the traces of the abstract model.
The intuition behind the trace inclusion based refinement is that a more abstract specification allows various
choices how system states may be reached while a more definite system description reduces those choices by
selecting the system traces corresponding to a more specific system implementation. This general notion of
reduction of system traces may be a design decision motivated by detected failures or simply motivated by
making the behaviour more deterministic and consequently needs to go along with various changes to the design.
This need for changes leads to  the following category of data refinement.

\subsubsection{Data refinement}
This refinement is defined by making an abstract data type more concrete in the refinement step.
A classical example is refining sets into the data type of lists which leads to making choices on order
and replication of elements: since lists are ordered, for example, the set $\{1,2,3\}$ could be
represented in the refinement by the list \texttt{[1,2,3]} but also \texttt{[3,2,1]} -- besides four
other orders. In addition, also \texttt{[1,1,2,3]} could be a possible refinement if replication
is deemed a targeted feature. As a consequence of this choice for refining the data, certain traces
are excluded in the refinement that could have been possible system traces under different choices for
order and replication.

\subsection{Choices, Failures, and Divergences}
To provide a closed representation of process behaviour, event based process calculi like CSP \cite{hoa:85,ros:98}
or CCS/Pi-calculus introduce choices between different events that are possible in the course of the execution of
a process. In the trace model it is simple to see the extension of those choices: there will be a number of different
continuations of a trace $t$, for example $t\tcon \langle a\rangle$ and $t\tcon \langle b\rangle$ represent the
potential continuation of a trace of events $t$ by either an event $a$ or an event $b$ if those two events are provided
by a choice at the point of execution of the process after trace $t$ has been executed. The syntax $\tcon$ represents
trace concatenation and $\langle a \rangle$ creates a trace consisting of just the event $a$.
In a purely event-based calculus, like CSP, CCS, or Pi, the internal system behaviour needs to be expressed
in the specification purely based on the reaction, since state information is not part of the model by contrast
to a model like the one used in the Isabelle Infrastructure framework (see previous subsection).
So to distinguish between different ways the specified system may react to the events that the environment offers,
CSP, for example, offers different kinds of choices
% from CSP notes
that correspond to two levels of non-determinism.
For any processes $P$ and $Q$ the process
\[ P \choi Q \]
is the so-called {\it external choice} between $P$ and $Q$. This
process is the process that offers the first events of $P$ and $Q$ to
the environment\footnote{The notion of environment means all other
  processes.} and then behaves either like $P$ or like $Q$ depending 
on the choice the environment made. Only if the environment offers
events that may be communicated by both processes there is a real
choice of the process.

The process
\[ P \sqcap Q \]
is the so-called {\it internal choice} or non-deterministic choice. In
contrast to the former, it expresses that the choice lies with the
system. That is, even if the environment offers only events that only
one of $P$ and $Q$ can communicate, the process may refuse to
communicate one of the offered events. Hence, it represents proper
non-determinism as the decision cannot be determined by the
environment.

%The difference is illustrated by the following examples.
%The process 
%\[ (a \rightarrow STOP) \choi (b \rightarrow STOP) \]
%has to communicate $a$ or $b$ if the environment offers one of them.
%The process
%\[ (a \rightarrow STOP) \sqcap (b \rightarrow STOP) \]
%may refuse both in that case. It only has to communicate if the
%environment offers both events simultaneously.

%the deterministic and nondeterministic or internal
%choice operators.
%In the deterministic choice, a process that has two alternatives, say $a$ and $b$, in the next step,
%must follow them if either of these events is offered by the environment. That is, if both events are possible
%there is really a choice for the process, but if only one of them is provided, the process cannot refuse.
%By contrast, the non-deterministic choice leaves this decision to the process. That is, this process could
%refuse to engage with an event, even if it was offered. Therefore, this process always has a choice.

In a refinement, non-deterministic choices can be replaced by deterministic choices: making a process
more deterministic moves it closer to an implementation. However, the traces model is not sufficient to
show the difference. In this model both processes above have the same set of traces and are thus semantically
equivalent:
$\text{traces}(P \choi Q) = \text{traces}(P \sqcap Q)$\,.

This motivates thinking about Failures.

\subsubsection{Failures}
In terms of modeling these different types of choices, it turns out that the pure trace model is not sufficient
because traces record only what processes can do but not what they must do or in other words what they can refuse
to do. In the example above, both processes have the trace with $a$ and $b$ included in their trace
set because for both this continuation is a possibility. However, the traces do not reflect that the process
using the non-deterministic choice could have stopped and refused to do anything in most cases.
Therefore, to integrate this additional view on choices, the CSP calculus offers a second semantics adding
refusals. Refusals are sets of events that a process can refuse at some point of execution, that is, for a certain
trace. Failures are the semantic model in which traces are combined with sets of refusals. In this semantics
our two processes above become distinguishable because $P \choi Q$ can only refuse what both $P$ and $Q$ can refuse
whereas $P \sqcap Q$ only must accept if neither $P$ nor $Q$ can refuse and thus in the failures model 
$\text{failures}(P \sqcap Q) \supset \text{failures}(P \choi Q)$ and consequently $P \sqcap Q \sqsubseteq_F P \choi Q$\,.

\subsubsection{Divergences and Termination}
Undefined behaviour may be represented by a system internal invisible action, for example, $\tau$ in CSP. 
Nontermination can then be defined by an infinite sequence of such $\tau$ actions. For those traces where
a program (or system) may diverge, the system does not communicate anything so there is no definite behaviour.
In order to arrive at a meaningful refinement order that augments failures and divergences, all system traces
are considered possible from the point of divergence onwards. Reducing non-termination leads to concretizing
a program corresponding to extending the range for which the program produces meaningful output. The most general
behaviour is represented by the process that diverges from the beginning. The divergences intuitively corresponds
more closely to a view of input-output program where classically termination is part of total correctness but is
less suitable for the wider system view we have here.

\subsubsection{Action refinement}
In reactive systems languages as well as human centric system formalisms, events or actions are a central modeling
entity. The terms ``action'' or ``event'' are mostly used synonymously but in some formalism, like statecharts, actions
and events are different entities -- one representing pre-requisites of transitions, the other their consequences.
Actions can be made more definite by combining them or making their occurrence dependent on conditions (or guards)
of the context. 

\subsection{Comparison of Isabelle's Infrastructure, State-based and Event-based Refinements}
\label{sec:isacomp}
The formal notion of refinement for Isabelle infrastructures with actors naturally enables all the different
notions of refinement because this refinement is simply based on an Isabelle type map and an inductive
definition of behaviour. Isabelle's datatypes are rich enough to support most data refinements.
Termination-based and non-deterministic refinement notions are typical for purely reactive systems where the
entire state is abstract and system behaviour is purely described by traces of events.
In our case, we might have the possibility of nontermination, for example by a loop in a state that may occur
infinitely often for all actions so that the state does not change.
Similarly, failures means that in a state some actions may not occur which can be modeled in the inductive
definition of the system behaviour.

A language like CSP is an explicit calculus for constructing specific specified communication patterns using
``standardized'' constructors.
However, using Higher Order Logic in Isabelle, we can simulate all choices, failures, etc, explicitly in the logic.
Even though we do not have standardized constructors, we can mimic each and any of the CSP constructors.
Moreover, we are not limited by a pure event-based view and can additionally simulate those constructors that CSP
cannot offer, like action refinement. The latter is possible because our actions are more abstract and
a vast portion of the action semantics is implemented in the additional preconditions of the corresponding
rule of the semantics for this action (see Section \ref{sec:inframod}) which provides fine grained tuning possibilities.

\section{Formal Definition of Refinement}
\label{sec:ref}
\subsection{Dependability Engineering}
As described initially, Dependability Engineering consists of two complimentary activities:
protection goal specification and attack (or, more generally, risk) analysis. These activities can be defined
informally by ``natural language'' descriptions or using semi-formal modeling languages. Attack
trees and Misuse Cases are commonly employed and recommended for this. For the design, UML diagrams
can be used. However, we take this general, informal or semi-formal methodology further in our Isabelle
Infrastructure framework that embeds attack trees in a fully formal way equipping it with a temporal logic
CTL semantics that supports finding attacks based on negating global policies.  Moreover, we adopt the system
specification refinement idea from classical software engineering whereby we overcome a crucial shortcoming
of refinement that has -- in our opinion -- impeded the success and the adoption of formal notions of
refinement: how to find the right refinement relation for a refinement step in a concrete application!

Finding the right refinement relation is in the security engineering process evidently provided
by the attacks found through the attack tree analysis. Since the attack properties as well as attack paths
are an outcome of the adequacy translation of the attack trees, these properties (in which the global security
property is violated) can be used. We want to avoid them. Since refinement is provided by trace sets, we do
this simply by excluding the attack traces and thereby we direct the refinement. Whether it is a data refinement,
action refinement of another form of refinement (see Section \ref{sec:refcomp}) may vary but all forms are
possible. We will illustrate the most relevant ones by our case study, after introducing the formal notion
of Dependability Refinement in the remainder of this section.

\subsection{Formal Definition of Refinement}
\label{sec:refform}
%\TODO{This section is just copied over from arxive paper - adapt}
Intuitively, a refinement changes some aspect of the type of
the state, for example, replaces a data type by a richer datatype or
restricts the behaviour of the actors. The former is expressed directly 
by a mapping of datatypes, the latter is incorporated into the state
transition relation of the Kripke structure that corresponds to the 
transformed model.
%This relationship between the Kripke structure can be better understood
%visually (see Figure \ref{fig:modtrans})
In other words, we can encode a refinement within our framework
as a relation on Kripke structures that is parameterized additionally by
a function that maps the refined type to the abstract type.
The direction ``from refined to abstract'' of this type mapping may seem 
curiously counter-intuitive. However, the actual refinement is given by the 
refinement that uses this function as an input. The refinement
then refines an abstract to a more concrete system specification. 
The additional layer for the refinement 
%(the red box in Figure \ref{fig:theorystruc})
can be formalized in Isabelle as a
%second\footnote{The first refinement 
%relation in this framework is on attack trees summarized in Section \ref{sec:at}.} 
refinement relation 
$\sqsubseteq_{\mathcal{E}}$. 
The relation \texttt{refinement} is typed as a relation over triples --
a function from a threefold Cartesian product to \texttt{bool}, the 
type containing true and false only.  
The type variables $\sigma$ and $\sigma'$ input to the type constructor 
\texttt{Kripke} represent the abstract state type and the concrete state type. 
Consequently, the middle element of the triples selected by the relation 
\texttt{refinement} is a function of type $\sigma' \Rightarrow \sigma$ 
mapping elements of the refined state to the abstract state.
The expression in quotation marks after the type is again the
infix syntax in Isabelle that allows the definition of mathematical notation
instead of writing \texttt{refinement} in prefix manner. This nicer infix
syntax is already used in the actual definition.
Finally, the arrow \texttt{\ttImp} is the implication of Isabelle's  meta-logic
while $\ttimp$ is the one of the {\it object} logic HOL. 
They are logically equivalent but of different
types: within a HOL formula $P$, for example, as below $\forall x. P \ttimp Q$, only the implication $\ttimp$
can be used.
\begin{ttbox}
 refinement :: (\ttsigma kripke \tttimes (\ttsigma' \ttfun \ttsigma) \tttimes \ttsigma' kripke)
                \ttfun bool ("_ \ttmref{(\_)} _")
  K \ttmeref K' \ttequiv \ttforall s' \ttin states K'. \ttforall s \ttin init K'. 
                    s \ttrelstar{\sigma'} s' \ttimp \ttecal(s) \ttin init K
                    \ttand \ttecal(s) \ttrelstar{\sigma} \ttecal(s')
\end{ttbox}
The definition of \texttt{K \ttmeref\, K'} states that for any state $s'$ 
of the refined Kripke structure that can be reached by the state transition
in zero or more steps from an initial state $s$ of the refined Kripke 
structure, the mapping ${\mathcal E}$ from the refined to the abstract 
model's state must preserve this reachability, i.e., the image of
$s$ must also be an initial state and from there the image of $s'$
under ${\mathcal E}$ must be reached with $0$ or $n$ steps.

\subsection{Property Preserving System Refinement}
A first direct consequence of this definition is the following lemma
where the operator \texttt{$\ttimg$} in \texttt{\ttecal\ttimg(init K')}
represents function image, that is the set, $\{\ttecal(x). x \in \texttt{init K'}\} $.
\begin{ttbox}
{\bf{lemma}} init_ref: K \ttmeref K' \ttImp \ttecal\ttimg(init K') \ttsubseteq init K
\end{ttbox}
A more prominent consequence of the definition of refinement 
is that of property preservation. Here, we show that refinement preserves the
CTL property of ${\sf EF} s$ which means that a reachability property true in the
refined  model \texttt{K'} %has also been 
is already true in the abstract model.
A state set $s'$ represents a property %since we use 
in the predicate transformer view of properties as sets of states. 
The additional condition on initial states ensures that we cannot ``forget'' them. 
%initial states.
%The theorem states that, if the 
%property \texttt{{\sf EF} s'} is true in the concrete model its  has been true 
%can be reached from the initial state in 
%\texttt{K'} then it is also possible  to reach the image of this property
% in the abstract Kripke structure \texttt{K}.
\begin{ttbox}
{\bf{theorem}} prop_pres: 
 K \ttmeref K'  \ttImp init K \ttsubseteq \ttecal\ttimg(init K') \ttImp
  \ttforall s' \ttin Pow(states K').
    K' \ttvdash {\sf EF} s' \ttimp K \ttvdash {\sf EF} (\ttecal\ttimg(s'))
\end{ttbox}
It is remarkable, that our definition of refinement by Kripke 
structure refinement entails property preservation and makes it possible 
to prove this as a theorem in Isabelle once for all, i.e., as a meta-theorem.
However, this is due to the fact that our generic definition of state transition
allows to explicitly formalize such sophisticated concepts like reachability.
For practical purposes, however, the proof obligation of showing that
a specific refinement is in fact a refinement is rather complex
justly because of the explicit use of the transitive closure of the state
transition relation.
In most cases, the refinement will be simpler. Therefore, we offer
additional help by the following theorem that uses a stronger characterization
of Kripke structure refinement and shows that our refinement follows
from this.
\begin{ttbox}
{\bf{theorem}} strong_mt: 
 \ttecal\ttimg(init K') \ttsubseteq init K \ttand s \ttrel{\sigma'} s' \ttimp \ttecal(s) \ttrel{\sigma} \ttecal(s') 
  \ttImp K \ttmeref K'
\end{ttbox}
This simpler characterization is in fact a stronger one: we could have $s \ttrel{\sigma'} s'$ 
in the refined Kripke structure \texttt{K'} and $\neg(\ttecal(s) \ttrel{\sigma} \ttecal(s'))$
but neither $s$ nor $s'$ are reachable from initial states in \texttt{K'}.
For cases, where we want to have the simpler one-step proviso but still need 
reachability we provide a slightly weaker version of \texttt{strong\_mt}.
\begin{ttbox}
{\bf{theorem}} strong_mt':  
 \ttecal\ttimg(init K') \ttsubseteq init K \ttand (\ttexists s0 \ttin init K'. s0  \ttrelIstar s)
  \ttand s \ttrel{\sigma'} s' \ttimp \ttecal(s) \ttrel{\sigma} \ttecal(s') \ttImp K \ttmeref K'
\end{ttbox}
The idea of property preservation coincides with the classical idea of
trace refinement as defined for process algebra semantics like CSP's. In these semantics,
the behaviour of a system is defined as the set of its traces, which are all possible sequences
of events reflecting each moment of system execution.
Consequently, a system S' is a refinement of another system S if the semantics of S' is a subset
of the traces of the former one.
We adopt this idea in principle, but extend it substantially: our notion additionally incorporates
structural refinement. Since we include a state map 
\texttt{\ttsigma'\ttfun \ttsigma} in our refinement map, we additionally
allow structural refinement: the state map generalizes the basic idea of
trace refinement by traces corresponding to each other but allows additionally
an exchange of data types thus allowing exchange of the structure of the system. 
As we see in the application to the case study in Section \ref{sec:corref}, the refinement steps may
sometimes just specialize the traces: in this case the state map 
\texttt{\ttsigma'\ttfun \ttsigma} is just identity. 

In addition, we also have a simple implicit version of {\it action refinement}. In an
action refinement, traces may be refined by combining consecutive system events
into atomic events thereby reducing traces.
We can observe this kind of refinement in the second refinement step
%This will be detailed in the second refinement step
of CWA considered in Section \ref{sec:secref}.

\subsection{Formal Definition of RR-Cycle}
\label{sec:rrcycfor}
Given the closed formal definition of refinement provided in the previous section, we
can now formalize the RR-Cycle by the following predicate.
\begin{ttbox}
  RR_cycle K K' s :: \ttsigma kripke \ttfun \ttsigma' kripke
                     \ttfun \ttsigma' set \ttfun bool
  RR_cycle K K' s \ttequiv
  \ttexists \ttecal :: \ttsigma' \ttfun \ttsigma. K \ttvdash {\sf EF}(\ttecal `s) \ttand K \ttmeref K'
  \ttimp \ttneg(K' \ttvdash {\sf EF} s)
\end{ttbox}
This predicate encompasses that the Kripke \texttt{K'} is a dependable system refinement of the Kripke
structure \texttt{K} if there exists a data type map \texttt{\ttecal} such that the refined Kripke structure
\texttt{K'} avoids the attack.

The RR-cycle is an iterative process. Since refinement is a transitive relation, the individual refinement
steps found in each iteration line up to one final refinement step. 
The formal predicate \texttt{RR\_cycle} constitutes the termination condition for the cycle for the specified
security goal \texttt{s} and for the lined up individual refinement steps. Only when a sufficient number
of interleaved refinement and attack tree iterations have eradicated all attacks on the security goal \texttt{s}
will the system specification be secure and this is verified by \texttt{RR\_cycle}. 

\section{Application Example Corona-Virus Warning App}
\label{sec:cwa}
The German Chancellor Angela Merkel has strongly supported the publication of
the mobile phone Corona-virus warning app (CWA; \cite{cwa:github}) by publicly proclaiming that
``this App deserves your trust'' \cite{bundes:20}. Many millions of mobile phone users
in Germany have downloaded the app with 6 million on the first day.
CWA is one amongst many similar applications that aim at the very important goal
to ``break infection chains'' by providing timely information to users of whether they
have been in close proximity to a person who tested positive for COVID-19.

%The CWA
%has taken a long time to develop being published only on
%16th June 2020. It
%was a quite costly project but this was mainly due to the management of
%Telekom and SAP being in the driving seat. But the app has been designed with great
The app was designed with great
attention on privacy: a distributed architecture \cite{cwa:arch} has been adopted %after a long and
%heated debate with supporters of a central architecture. The distributed architecture is
that is
based on a very clever application design whereby clients broadcast
highly anonymized identifiers (ids)
%so called ``Ephemeral Ids''
via Bluetooth and store those ids of people in close proximity.
%highly anonymized so called ``Ephemeral IDs'' at physical locations via the
%Bluetooth Low Energy Beacon protocol.
%The app saves those ids of people in close proximity.
Infected persons report their infection by uploading their ids to a central server, providing all clients the means to assess exposure risk locally,
hence, stored contact data has never to be shared.
%Infected persons report their infection to a central server,
%which, in turn, provides all clients the means to reconstruct the ids
%used by reported cases over the last 14 days.
%
%When at a later date an infected person reports his infection to a central server,
%all clients are provided the means to reconstruct the ids used by this particular
%person's device over the last 14 days.
%Therefore, exposure risks can be evaluated locally and stored contact data has never to be shared.
%
%, the unique root ID is published
%and in the daily check all mobile phones connecting to the central server download
%the root IDs of infected people. Since the Ephemeral IDs can be mapped to the root ID
%all Ephemeral IDs that have been saved over the last 14 days allow users' phones to
%regularly check whether their user has been exposed to an infected person and issue
%a warning to the user and recommendation to contact health authorities.
%The warning issued by the Corona warning app entitles to
%having a Corona test done (which at the time of writing is not normally possible).

\subsection{DP-3T and PEPP-PT}
\label{sec:history}
We are mainly concerned with the architecture and protocols proposed by the
DP-3T (\textit{Decentralized Privacy-Preserving Proximity Tracing}) project \cite{dp3t:github}.
The main reason to focus on this particular family of protocols is the
\textit{Exposure Notification Framework} (ENF), jointly published by Apple and Google \cite{enf:proj}, that adopts
%the main
core principles of the DP-3T proposal. This API is not only used in 
CWA but has the potential of being widely adopted in future
app developments that might emerge due to the reach of players like Apple and Google.

There are, however, competing architectures noteworthy, namely protocols developed under the
roof of the \textit{Pan-European Privacy-Preserving Proximity Tracing} project (PEPP-PT) \cite{pepppt:github}, e.g.
PEPP-PT-ROBERT \cite{pepppt:ROBERT},
that might be characterized as centralized architectures.

Neither DP-3T nor PEPP-PT are synonymous for just one single protocol. Each project endorses
different protocols with unique properties in terms of privacy and data protection.
%Yet, on a higher level of abstraction, it seems feasible to distinguish two basic architectures:
%protocols as endorsed by PEPP-PT might be characterized as centralized architectures whereas
%DP-3T-inspired protocols follow a (more) decentralized approach.\footnote{%
%  DP-3T involves a central backend server. It is decentralized with regard to
%  the collection and evaluation of contact information:
%  In centralized architectures the server provides a risk scoring services, whereas decentralized
%  approaches rely on local risk assessment and, thus, do not need to share contact information with
%  the backend.}

There is a variety of noteworthy privacy and security issues. %that deserve formal consideration (and verification).
The debate among advocates of centralized architectures and those in favor of a decentralized approach in particular yields a lot of interesting material detailing different attacks and possible mitigation strategies: \cite{pepppt:dp3tana}, \cite{pepppt:dp3tresp}, \cite{dp3t:psre}.

In terms of attack scenarios, we focus on, what might be classified as deanonymisation attacks: Tracking a device (see \cite{dp3t:psre}[p9], \cite{pepppt:dp3tana}[p8]) and identifying infected individuals (see \cite{dp3t:psre}[p5]\cite{pepppt:dp3tana}[p9]).

\subsubsection{Basic DP-3T architecture}
Upon installation, the app generates secret daily seeds to derive so-called \textit{Ephemeral Ids}
(EphIDs) from them. EphIDs are generated locally with cryptographic methods and cannot be connected
to one another but only reconstructed from the secret seed they were derived from.

During normal operation each client broadcasts its EphIDs via Bluetooth whilst scanning for
EphIDs broadcasted by other devices in the vicinity. Collected EphIDs are stored locally along
with associated meta-data such as signal attenuation and date. In DP-3T the contact information
gathered is never shared but only evaluated locally.

If patients test positive for the Corona-virus, they are entitled to upload specific data to a
central backend server. This data is aggregated by the backend server and redistributed to all
clients regularly to provide the means for local risk scoring, i.\,e., determining whether collected
EphIDs match those broadcasted by now-confirmed Corona-virus patients during the last, e.\,g., 14, 
days.

\begin{figure*}%[htb]
%\vspace{-.5cm}
  \begin{center}
    \includegraphics[width=5in]{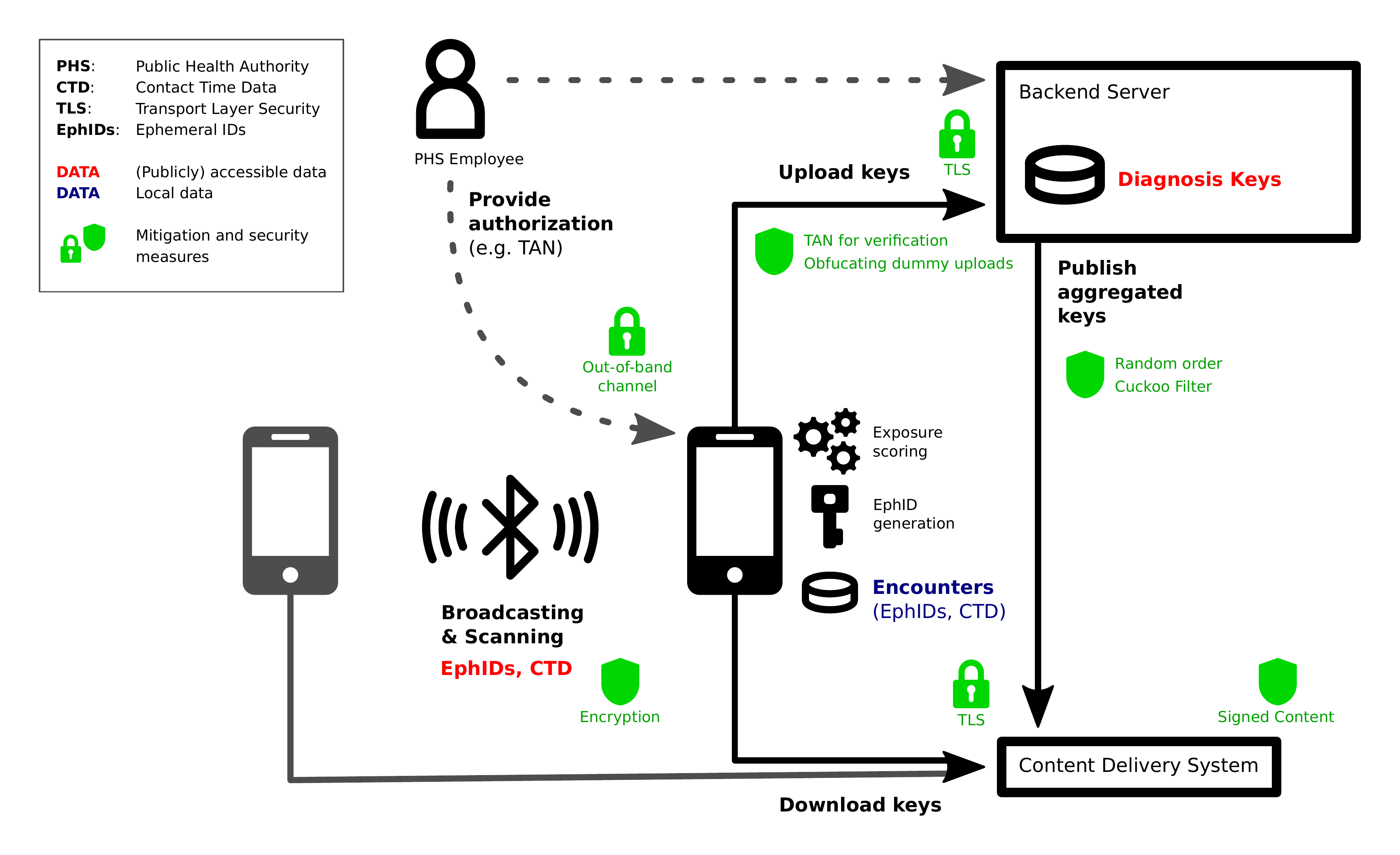}
  \end{center}
%\vspace{-.5cm}
  \caption{Decentralized privacy-preserving proximity tracing protocol of CWA}
  \label{fig:dp3tprot}
%\vspace{-.5cm}
\end{figure*}

In the most simple (and insecure) protocol proposed by DP-3T %\cite{dp3t:wp}[p14ff]
this basically translates into publishing the daily seeds used to derive EphIDs from.
The protocol implemented by ENF and,
hence, CWA adopts this low-cost design \cite{cwa:arch}. DP-3T proposes two other, more sophisticated protocols that improve
privacy and data protection properties to different degrees but are more costly to implement.
Figure \ref{fig:dp3tprot} illustrates the basic system architecture
along with some of the mitigation measures either implemented in CWA or proposed by DP-3T.

\subsection{Instantiation of Framework}
The formal model of CWA uses the Isabelle Infrastructure framework instantiating it 
by reusing its concept of {\it actors} for users and smartphones whereby locations correspond
to physical locations. The Ephemeral Ids, their sending and change is added to Infrastructures
by slightly adapting the basic state type of infrastructure graphs and accordingly the semantic rules
for the actions move, get, and put. The details of the newly adapted Infrastructure are
presented in Section \ref{sec:model}.
Technically, an Isabelle theory file \texttt{Infrastructure.thy} builds on top of the theories for Kripke 
structures and CTL (\texttt{MC.thy}), attack trees (\texttt{AT.thy}), and security refinement 
(\texttt{Refinement.thy}). Thus all these concepts can be used to specify the formal model
for CWA, express relevant and interesting properties, and conduct
interactive proofs (with the full support of the powerful and highly automated proof support
of Isabelle).
%The CWA theory itself is an adaptation of the Infrastructure theory of the 
%Isabelle Infrastructure framework and reuses (or slightly adapts) existing concepts. 
%In the remainder of this paper, we introduce the model that we conceived for . 
All Isabelle sources are available online \cite{kam:20gitsc}.

\section{Modeling and analyzing CWA}
\label{sec:model}
\subsection{Infrastructures, Policies, and Actors}
\label{sec:inframod}
% Adapted from CSF paper
The Isabelle Infrastructure framework supports the representation of infrastructures 
as graphs with actors and policies attached to nodes. These infrastructures 
are the {\it states} of the Kripke structure. % for the attack trees.

The transition between states is triggered by non-parameterized
actions \texttt{get}, \texttt{move}, and \texttt{put} 
executed by actors. 
Actors are given by an abstract type \texttt{actor} and a function 
\texttt{Actor} that creates elements of that type from identities 
(of type \texttt{string} written \texttt{''s''} in Isabelle). 
Actors are contained in an infrastructure graph type \texttt{igraph}
constructed by its constructor \texttt{Lgraph}. 
\begin{ttbox}
{\bf datatype} igraph = 
 Lgraph (location \tttimes location)set 
  location \ttfun identity set
  identity \ttfun (string set \tttimes string set \tttimes efid)  
  location \ttfun string \tttimes (dlm \tttimes data) set
  location \ttfun efid set
  identity \ttfun location \ttfun (identity \tttimes efid) set
\end{ttbox}
In the current application of the framework to the CWA case study, 
this graph contains a set of location pairs representing the %structure 
topology of the infrastructure
as a graph of nodes and a function %\footnote{ We use the common $\lambda$-abstraction, 
%e.g. \texttt{\ttlam x. True}, to define functions with parameters, here the 
%function returning True for any input $x$.} 
that assigns a set of actor identities to each node (location) in the graph.
The third component of an \texttt{igraph} assigns the credentials to each actor: 
a triple-valued function whose first range component is a set describing the credentials 
in the possession of an actor and the second component 
is a set defining the roles the actor can take on; 
most prominently the third component is the \texttt{efid} assigned to the actor. This is initially
just a natural number but will be refined to actually represent lists of Ephemeral Ids later
when refining the specification.
\begin{ttbox}
{\bf datatype} efid = Efid nat
\end{ttbox}
The fourth component of the type \texttt{igraph}
assigns security labeled data to locations, a feature not used in the
current application.
The second to last component assigns the set of efids of all currently present smart phones
to each location of the graph. The last component finally denotes the knowledge set of each
actor for each location: a set of pairs of actors and potential ids.

Corresponding projection functions for each of the components of an 
infrastructure graph are provided; they are named \texttt{gra} for the actual 
set of pairs of locations, \texttt{agra} for the actor map, \texttt{cgra} for 
the credentials, \texttt{lgra} for the %state of a location and the 
data at that location, \texttt{egra} for the assignment of current efids to locations,
and \texttt{kgra} for the knowledge set for each actor for each location.

To start the CWA case study, we provide an example scenario: the initial values for the
\texttt{igraph} components use two locations \texttt{pub} and \texttt{shop} to define the
following constants (we omit the data map component \texttt{ex\_locs}).
The map \texttt{ex\_loc\_ass} is a function, defined using the $\lambda$-calculus of Isabelle,
mapping each location to the set of actor identities currently at this location. 
\begin{ttbox}
 ex_loc_ass \ttequiv
  (\ttlam x. if x = pub then \{''Alice'', ''Bob'', ''Eve''\}  
      else (if x = shop then \{''Charly'', ''David''\} 
      else \{\}))
\end{ttbox}
The component \texttt{ex\_creds} is a function assigning actor identities to triplets
of which the most important part is the third element that contains the actors' ephemeral ids.
\begin{ttbox}
 ex_creds \ttequiv
  (\ttlam x. if x = ''Alice'' then Efid 1 else 
     (if x = ''Bob'' then Efid 2 else 
     (if x = ''Charly'' then Efid 3 else
     (if x = ''David'' then Efid 4 else
     (if x = ''Eve'' then Efid 5
      else Efid 0)))))
\end{ttbox}
The ephemeral ids of all actors present at a location are collected in the
component \texttt{ex\_efids} that maps each location to the set of these efids.
\begin{ttbox}
 ex_efids \ttequiv
  (\ttlam x. if x = pub then \{Efid 1, Efid 2, Efid 5\}
   else (if x = shop then \{Efid 3, Efid 4\} else \{\}))
\end{ttbox}
The final component \texttt{ex\_knos} assigns actors to functions mapping each location
to sets of pairs of actor identities and efids: these sets of pairs represent what can be inferred
from observing the present actors and all efids at a location; simply all possible combinations
of those. Initially, however the empty set of pairs is assigned to all locations because nothing
is known yet. 
\begin{ttbox}
 ex_knos \ttequiv
 (\ttlam x. (if x = Actor ''Eve'' then (\ttlam l. \{\})
   else (\ttlam l. \{\})
\end{ttbox}  
These components are wrapped up into the following \texttt{igraph}.
\begin{ttbox}
 ex_graph \ttequiv
  Lgraph \{(pub, shop)\}
         ex_loc_ass ex_creds ex_locs ex_efids ex_knos
\end{ttbox}  
Infrastructures are given by the following datatype that 
contains an infrastructure graph of type \texttt{igraph} 
and a policy given by a function that assigns local policies over a graph to
all locations of the graph. 
\begin{ttbox}
 {\bf{datatype}} infrastructure =
 Infrastructure igraph
               [igraph, location] \ttfun policy set
\end{ttbox}
There are projection functions \texttt{graphI} and \texttt{delta} when applied
to an infrastructure return the graph and the local policies, respectively.
The function \texttt{local\_policies} gives the policy for each location \texttt{x}
over an infrastructure graph \texttt{G} as a pair: the first element of this pair is 
a function specifying the actors \texttt{y} that are entitled to perform the actions 
specified in the set which is the second element of that pair.
The local policies definition for CWA, simply permits all actions to
all actors in both locations.
\begin{ttbox}
 local_policies G \ttequiv
  (\ttlam x. if x = pub then
       \{(\ttlam y. True, \{get,move,put\})\}
         else (if x = shop
               then \{(\ttlam y. True, \{get,move,put\})\}
               else \{\}))
\end{ttbox}  
For CWA, the initial infrastructure contains the graph \texttt{ex\_graph}
with its two locations pub and shop and is then wrapped up with the local policies
into the infrastructure \texttt{Corona\_scenario} 
that represents the ``initial'' state for the Kripke structure.
\begin{ttbox}
 Corona_scenario \ttequiv
  Infrastructure ex_graph local_policies
\end{ttbox}

\subsection{Policies, privacy, and behaviour}
\label{sec:}
Policies specify the expected behaviour of actors of an infrastructure. 
They are given by pairs of predicates (conditions) and sets of (enabled) actions.
%\begin{ttbox}
%{\bf type}_{\bf{synonym}} policy = ((actor \ttfun bool) \tttimes action set)
%\end{ttbox}
%The behaviour of actors is 
They are defined by the \texttt{enables} predicate:
an actor \texttt{h} is enabled to perform an action \texttt{a} 
in infrastructure \texttt{I}, at location \texttt{l}
if there exists a pair \texttt{(p,e)} in the local policy of \texttt{l}
(\texttt{delta I l} projects to the local policy) such that the action 
\texttt{a} is a member of the action set \texttt{e} and the policy 
predicate \texttt{p} holds for actor \texttt{h}.
\begin{ttbox}
enables I l h a \ttequiv \ttexists (p,e) \ttin delta I l. a \ttin e \ttand p h
\end{ttbox} 

The privacy protection goal is to avoid deanonymization. That is, an attacker should not be able to
disambiguate the set of pairs of real ids and EphIDs. This is abstractly expressed in the predicate
identifiable. The clue is to filter the set \texttt{A} of all pairs of real ids and
ephemeral ids \texttt{eid} for the given \texttt{eid}. If this set happens to have only one element,
that is, there is only one pair featuring this specific \texttt{eid}, then the real id of that efid
must be the first component, the identity \texttt{Id}.
\begin{ttbox}
 identifiable eid A \ttequiv
  is_singleton\{(Id,Eid). (Id, Eid) \ttin A \ttand Eid = eid\}
\end{ttbox}
The predicate identifiable is used to express the global policy `Eve cannot deanonymize an Ephemeral
Id \texttt{eid} using the gathered knowledge'.
%{\bf fixes} global_policy::[infrastructure, identity] \ttfun bool
%{\bf defines}  
\begin{ttbox}
 global_policy I eid \ttequiv
 \ttforall L \ttsubseteq nodes(graphI I). 
 \ttneg(identifiable eid
     (\ttbigcap (kgra(graphI I) (Actor ''Eve'')` L)
      - \{(x,y). x = ''Eve''\}))
\end{ttbox}
Gathering the ``observations'' of the actor Eve over all locations in the above policy formula
is achieved by combining all pairs of ids and efids that Eve could infer on her movements through
different locations, then building the intersection over those sets to condense the 
information and finally substracting all pairs in which Eve's own identity appears as first element.
The resulting set of pairs should not allow identification of any \texttt{eid}, that is, it must
not be singleton for the global policy to hold, and thus, to protect privacy of users.
\subsection{Infrastructure state transition}
\label{sec:statetrans}
The state transition relation is defined as a syntactic infix notation 
\texttt{I \ttrelI\, I'}  and denotes that infrastructures 
\texttt{I} and \texttt{I'} are in this relation.
%Rules for the put, get, move, process and delete actions exist.
To give an impression of this definition, we show first the rule defining the state transition for
the action get. Initially, this rule expresses that
an actor that resides at a location \texttt{l} in graph \texttt{G} (denoted as ``\texttt{a \ttatI\ l}'')
and is enabled by the local policy in this location to ``get'', can
combine  all ids at the current location (contained in \texttt{egra G l}) with
all actors at the current location (contained in \texttt{agra G l}) and add this
set of pairs to his knowledge set \texttt{kgra G} using the function update \texttt{f(l := n)}
redefining the function \texttt{f} for the input \texttt{l} to have now the new value \texttt{n}.
\begin{ttbox}
  {\bf{get}}: G = graphI I \ttImp a \ttatI l \ttImp
  enables I l (Actor a) get \ttImp 
  I' = Infrastructure 
      (Lgraph (gra G)(agra G)(cgra G)(lgra G)(egra G)
      ((kgra G)((Actor a) := ((kgra G (Actor a))(l:=
        \{(x,y). x \ttin agra G l \ttand y \ttin egra G l\})))))
      (delta I)
  \ttImp I \ttrel{i} I' 
\end{ttbox}
Another interesting rule for the state transition is the one for \texttt{move} whose structure
resembles the previous one.
\begin{ttbox}
 {\bf{move}}: G = graphI I \ttImp a \ttatI l \ttImp
   a \ttin actors_graph(graphI I) \ttImp l \ttin nodes G \ttImp
   l' \ttin nodes G \ttImp enables I l' (Actor h) move \ttImp
   I' = Infrastructure
          (move_graph_a a l l' (graphI I))(delta I)
 \ttImp I \ttrelI I' 
\end{ttbox}
The semantics of this rule is implemented in the function \texttt{move\_graph\_a}
that adapts the infrastructure state so that the moving actor \texttt{a} is now
associated to the target location \texttt{l'} in the actor map \texttt{agra} and not
any more at the previous location \texttt{l} and also the association of \texttt{efids}
is updated accordingly.
\begin{ttbox}
move_graph_a n l l' g \ttequiv
 Lgraph (gra g) 
  (if n \ttin ((agra g) l) \ttand  n \ttnin ((agra g) l') then 
    ((agra g)(l := (agra g l) - \{n\}))
     (l' := (insert n (agra g l'))) else (agra g))
  (cgra g)(lgra g)
  (if n \ttin ((agra g) l) \ttand  n \ttnin ((agra g) l') then
    ((egra g)(l := (egra g l) - \{cgra g n\}))
     (l' := (insert cgra g n)(egra g l')))
      else egra g)
  (kgra g)
\end{ttbox}  
Based on this state transition and the above defined \texttt{Corona\_scenario}, we define the first Kripke structure.
\begin{ttbox}
 corona_Kripke \ttequiv Kripke \{I. Corona_scenario \ttrelIstar I\}
                       \; \{Corona_scenario\}
\end{ttbox}

\subsection{Attack analysis}
\label{sec:ana}
For the analysis of attacks, we negate the security property that we want to achieve,
usually expressed as the global policy.

Since we consider a predicate transformer semantics, we use
sets of states to represent properties. 
%For example, the attack property
The invalidated global policy is given by the set \texttt{scorona}.
\begin{ttbox}
 scorona \ttequiv \{x. \ttexists n. \ttneg global_policy x (Efid n)\}
\end{ttbox}
The attack we are interested in is to see whether for the scenario
\begin{ttbox}
 Kripke\_scenario \ttequiv  Infrastructure ex_graph
                                  \, local_policies 
\end{ttbox}
from the initial state \texttt{Icorona \ttequiv \{corona\_scenario\}},
the critical state \texttt{scorona} can be reached,
that is, is there a valid attack \texttt{(Icorona,scorona)}?

As an initial step, we derive a valid and-attack for the Kripke structure \texttt{corona\_Kripke}
using the attack tree proof calculus.
\begin{ttbox}
  \ttvdash [\ttcalN{(Icorona,Corona)},\ttcalN{(Corona,Corona1)},\ttcalN{(Corona1,Corona2)},
           \ttcalN{(Corona2,Corona3)},\ttcalN{(Corona3,scorona)}]\ttattand{\texttt{(Icorona,scorona)}}
\end{ttbox}
The properties (and thus sets) \texttt{Corona, Corona1, Corona2, Corona3} are intermediate states given
by \texttt{Bob} moving to shop and \texttt{Eve} following him while collecting the Ephemeral Ids in
each location. This collected information enables Eve to identify Bob's Ephemeral Id.

From these preparations, we can exhibit that an attack is possible using the attack tree calculus \cite{kam:18b}.
\begin{ttbox}
 corona_Kripke \ttvdash {\sf EF} scorona
\end{ttbox}
To this end, the Correctness theorem \texttt{AT\_EF} is simply applied to 
immediately prove the above CTL statement. This application of the meta-theorem 
of Correctness of attack trees saves us proving the CTL formula tediously 
by exploring the state space in Isabelle proofs. Alternatively, we could use 
the generated code for the function \texttt{is\_attack\_tree} in Scala 
(see \cite{kam:18b}) to check that a refined attack of the above is valid.

\section{Application of Refinement to CWA}
\label{sec:cwaref}
\subsection{Refining the Specification}
\label{sec:corref}
Clearly, fixed Ephemeral Ids are not really ephemeral. The model presented
in Section \ref{sec:model} has deliberately been designed abstractly to allow focusing on
the basic system architecture and finding an initial deanonymization attack.
We now introduce ``proper'' Ephemeral Ids and show how the system datatype can be refined to a system
that uses those instead of the fixed ones.

For the DP-3T Ephemeral Ids \cite{dp3t:wp}, for each day $t$ a seed $SK_t$
is used to generate a list of length $n = 24 * 60 / L$, where $L$ is the duration for which
the Ephemeral Ids are posted by the smart phone
\begin{ttbox}
 EphID1 || ... || EphIDn =
                    PRG(PRF(SKt,``broadcast key''))
\end{ttbox} 
``where PRF is a pseudo-random function (e.g., HMAC-SHA256), ``broadcast key'' is a
fixed, public string, and PRG is a pseudorandom generator (e.g. AES in counter mode)'' \cite{dp3t:wp}.

From a cryptographic point of view, the crucial properties of the Ephemeral Ids are that
they are purely random, therefore, they cannot be guessed, but at the same time if -- after
the actual encounter between sender and receiver -- the seed $SK_t$ is published, it is feasible
to relate any of the \texttt{EphIDi} to $SK_t$  for all \texttt{i} $\in \{1, \ldots, n\}$.
For a formalization of this crucial cryptographic property in Isabelle it suffices to define a new type
of list of Ephemeral Ids \texttt{efidlist} containing the root $SK_t$ (the first \texttt{efid}),
a current \texttt{efid} indicated by a list pointer of type \texttt{nat}, and the actual list of
\texttt{efid}s given as a function from natural numbers to \texttt{efid}s.
\begin{ttbox}
{\bf{datatype}} efidlist = Efids "efid" "nat" "nat \ttfun efid"
\end{ttbox}
We define functions for this datatype: \texttt{efidsroot} returning the first of the three
constituents in an \texttt{efidlist} (the root $SK_t$); \texttt{efids\_index} giving the second
component, the index of the current \texttt{efid}; \texttt{efids\_inc\_ind} applied to an
\texttt{efidlist} increments the index; \texttt{efids\_cur} returning the current \texttt{efid}
from the list and \texttt{efids\_list} for the entire list (the third component).

The first step of refinement replaces the simple \texttt{efid} in the type \texttt{igraph} of the
infrastructure type by the new type \texttt{efidlist}. Note, that in the new datatype \texttt{igraph}
this change affects only the third component, the credentials, to become
\begin{ttbox}
   identity \ttfun efidlist
\end{ttbox}
%The last two components, the set of currently present efids,
%and the knowledge set, remain the same and still operate on the simple type \texttt{efid}.
%\begin{ttbox}
%{\bf datatype} igraph =
%         Lgraph (location \tttimes location)set
%                 location \ttfun identity set
%
%                 location \ttfun string \tttimes (dlm \tttimes data) set
%                 location \ttfun efid set
%                 actor \ttfun location \ttfun (identity \tttimes efid) set
%\end{ttbox}
The refined state transition relation implements the possibility of changing the Ephemeral Ids
by the rule for the action \texttt{put} that resembles very much the rule for \texttt{get}.
%\begin{ttbox}
% {\bf{put}}: G = graphI I \ttImp  a \ttatI l \ttImp enables I l (Actor a) put \ttImp
%      I' = Infrastructure (put_graph_efid A l (graphI I))(delta I)
% \ttImp I \ttrelI I'
%\end{ttbox}
The important change to the infrastructure state is implemented in the function \texttt{put\_graph\_efid} that
increases the current index in the \texttt{efidlist} in the credential component \texttt{cgra g n} for
the ``putting'' actor identity \texttt{n} and inserts the current \texttt{efid} from that credential
component into the \texttt{egra} component, the set of currently ``present'' Ephemeral Ids at the location
\texttt{l}.
\begin{ttbox}
 put_graph_efid n l g  \ttequiv
  Lgraph (gra g)(agra g)
   ((cgra g)(n :=  efids_inc_ind(cgra g n)))
   (lgra g)
   ((egra g)
     (l := insert (efids_cur(efids_inc_ind(cgra g n)))
                  ((egra g l) - \{efids_cur(cgra g n)\})))
   (kgra g)
\end{ttbox}
We can now apply the refinement by defining a datatype map from the refined infrastructure type
\texttt{InfrastructureOne.infrastructure} to the former one \texttt{Infrastructure.infrastructure}.
\begin{ttbox}
{\bf{definition}} refmap :: InfrastructureOne.infrastructure 
                \ttfun Infrastructure.infrastructure
{\bf{where}} refmap I =
  Infrastructure.Infrastructure 
    (Infrastructure.Lgraph
       (InfrastructureOne.gra (graphI I))
       (InfrastructureOne.agra (graphI I))
        (\ttlam h. repl_efr 
           (InfrastructureOne.cgra (graphI I)) h)
           (InfrastructureOne.lgra (graphI I))
(\ttlam l. \ttlam a. 
     efids_root (InfrastructureOne.cgra (graphI I) a)
     \ttimg(InfrastructureOne.agra (graphI I) l)
   (\ttlam a. \ttlam l.
     if (a \ttin actors_graph(graphI I) \ttand
         l \ttin nodes (graphI I))
     then (\ttlam (x,y).
     (x, efids_root(
         InfrastructureOne.cgra (graphI I)
              (anonymous_actor I y)))
        \ttimg(InfrastructureOne.kgra (graphI I)) a l)
     else {})))   
\end{ttbox}
The function \texttt{anonymous\_actor} maps an efid back to the actor it is assigned to
if that exists which can be expressed in Isabelle using the Hilbert-operator \texttt{SOME}.
\begin{ttbox}
  anonymous_actor I e = (SOME a :: identity.
    a \ttin actors_graph (InfrastructureOne.graphI I) \ttand
    e \ttin range(efids_list (cgra (graphI I) a)))
\end{ttbox}  
This refinemnent map is then plugged into the parameter \texttt{\ttecal} of the refinement operator allowing to
prove
\begin{ttbox}
 corona\_Kripke \ttref{\texttt{refmap}} corona\_KripkeO
\end{ttbox}    
where the latter is the refined Kripke structure.
The proof of this refinement that adds quite some structure requires a great deal of proving of
invariants over the datatype.

Surprisingly, we can still prove
\begin{ttbox}
 corona\_KripkeO \ttvdash EF scoronaO
\end{ttbox}
by using the same attack tree as in the abstract model: if Bob moves from pub to shop, he is
vulnerable to being identifiable as long as he does not change the current \texttt{efid}. So,
if Eve moves to the shop as well and performs a get before Bob does a put, then Eve's knowledge
set permits identifying Bob's current Ephemeral Id as his. Despite removing some attack paths,
some remain and thus the identification attack persists.

\subsection{Second Refinement Step}
\label{sec:secref}
The persistent attack can be abbreviated informally by the action sequence \texttt{[get,move,move,get]}
performed by actors Eve, Bob, Eve, and Eve again, respectively.
How can a second refinement step avoid that Eve
%acts out the last
can \texttt{get} identification information for Bob? Can we impose that after the
first \texttt{move} of Bob a \texttt{put} action, increasing Bob's efid and thus obfuscating his
id, must happen before Eve can do another \texttt{get}? 
A very simple remedy to impose this and thus exclude the persistent attack is to enforce a \texttt{put}
action after every \texttt{move} by an action refinement that binds these actions together.
That is, when actors \texttt{move} they simultaneously change their efid.
We can implement that change by a minimal update to the function \texttt{move\_graph\_a} (see Section \ref{sec:model})
by adding an increment (highlighted as \texttt{\bf efids\_inc\_ind} in the code snippet)
of the currently used Ephemeral Id before updating the \texttt{egra} component of the target location.
In other words, if Bob moves his \texttt{egra} component is changed automatically, that is, he
gets a new efid, to stay with the example.
\begin{ttbox}
move\_graph\_a n l l' g \ttequiv
 Lgraph (gra g) 
  (if n \ttin ((agra g) l) \ttand  n \ttnin ((agra g) l') then 
    ((agra g)(l := (agra g l) - \{n\}))
     (l' := (insert n (agra g l'))) else (agra g))
  (cgra g)(lgra g)
  (if n \ttin ((agra g) l) \ttand  n \ttnin ((agra g) l') then
    ((egra g)(l := (egra g l) - \{cgra g n\}))  
    (l' := insert (efids_cur({\bf efids\_inc\_ind}
          (cgra g n)))(egra g l')))
     else egra g)
  (kgra g)
\end{ttbox}  
This is an action refinement because the move action is changed. It is a refinement, since
any trace of the refined model can still be mapped to a trace in the more abstract model just
omitting a few steps (the refinement relation is defined using the reflexive transitive closure
\texttt{\ttrelIstar}):
The refinement map \texttt{refmapI} is trivial since the datatypes do not change.
Because of \texttt{\ttrelIstar} intermediate dangerous states in the abstract may be
swallowed in the refined model. For example, the state $s$, where Bob has moved but not yet
put the new efid, does not exist in the refined model any more.

\subsection{Finalizing RR-Cycle Analysis}
As we have seen in Section \ref{sec:rrcycfor} and as has been illustrated by the application example, the
RR-cycle is an iterative process. The formal predicate constitutes the termination condition for the cycle
for the specified security goal: in our case study this has been ``non-identifiability''.
Applying the formal characterization of the predicate \texttt{RR-cycle} to the case study after the two steps
detailed in this section, we see that the second refinement step still does not avoid the deanonymization attack.
Trying to verify the predicate \texttt{RR\_cycle} fails.
We find another attack possibility: actor Alice is on her own with the attacker in a location.
As a remedy, the final refinement steps ensure that action \texttt{move} is constrained by (a) no actor
can move to a location where there are less than three actors and (b) also no actor leaves a location
when there are not at least four actors.
Step (a) is achieved by refining the specification such that it is only permitted to move
if there are three or more actors at the destination location \texttt{l'}. Isabelle's mathematical library provides the
necessary arithmetic and set theory to express this constraint as \texttt{card (agra g l') \ttgeq\ 3}. The constraint
can be added as an additional precondition to the function \texttt{move\_graph\_a} for the field \texttt{agra}
(and similarly for the \texttt{egra} field, see previous Section \ref{sec:secref}).
Similarly, Step (b) can be achieved by refining the specification further to allow an actor to move if the source
destination \texttt{l} has at least three actors by adding \texttt{card (agra g l') \ttgeq\ 4} to the \texttt{agra}
field and similarly for \texttt{egra}.
Summarizing, these two steps lead to a third refinement step that does not change the datatypes
(as in the previous refinement step) but restricts the set of possible traces of state transition
relations as is manifested in the refined 
definition of \texttt{move\_graph\_a}. The additional conditions are highlighted.
\begin{ttbox}
move\_graph\_a n l l' g \ttequiv
 Lgraph (gra g) 
 (if n \ttin ((agra g) l) \ttand  n \ttnin ((agra g) l')
    {\bf \ttand card (agra g l') \ttgeq 3 \ttand card (agra g l) \ttgeq 4}
  then ((agra g)(l := (agra g l) - \{n\}))
      (l' := (insert n (agra g l'))) else (agra g))
  (cgra g)(lgra g)
  (if n \ttin ((agra g) l) \ttand  n \ttnin ((agra g) l') 
     {\bf \ttand card (agra g l') \ttgeq 3 \ttand card (agra g l) \ttgeq 4}
   then ((egra g)(l := (egra g l) - \{cgra g n\}))  
    (l' := insert (efids_cur({efids\_inc\_ind}
          (cgra g n)))(egra g l')))
   else egra g)
  (kgra g)
\end{ttbox}  
After this refinement, the \texttt{RR\_cycle} verification succeeds: we can finally prove
the following theorem.
%, and we thus know that the following is a theorem.
%Thus, the above theorem corresponds to the final verification of the termination condition \texttt{RR\_cycle}
%because it is equivalent to
\begin{ttbox}
  corona\_KripkeR \ttvdash
      \ttneg {\sf EF} \{x. \ttexists n. \ttneg global\_policy x (Efid n))\}
\end{ttbox}
The proved meta-theory for attack trees can be applied to make this theorem more intuitive:
the contraposition of the Correctness property grants that if there is no attack on
\texttt{(I,\ttneg f)}, then \texttt{{\sf EF} \ttneg f} does not hold in the Kripke structure.
This yields the theorem in the following more understandable form
since the \texttt{{\sf AG} f} statement corresponds to \texttt{\ttneg {{\sf EF} \ttneg f}}.
\begin{ttbox}
 corona\_KripkeR \ttvdash
      {\sf AG} \{x. \ttforall n. global\_policy x (Efid n)\}
\end{ttbox}
%The equivalence
%is proved as a meta-theorem for CTL in the Isabelle Infrastructure framework illustrating how the expressivity
%of Isabelle allows proving generic meta-theorems and then applying them to applications.

\section{Related Work}
\label{sec:related}

\subsection{Isabelle Insider and Infrastructure framework}
\label{sec:relisa}
A whole range of publications have documented the development of the Isabelle Insider framework.
The publications \cite{kp:13,kp:14,kp:16} first define the fundamental notions of insiderness, policies,
and behaviour showing how these concepts are able to express the classical insider threat patterns
identified in the seminal CERT guide on insider threats \cite{cmt:12}.
This Isabelle Insider framework has been applied to auction protocols \cite{kkp:16,kkp:16a} illustrating
that the Insider framework can embed the inductive approach to protocol verification \cite{pau:98}.
An Airplane
case study \cite{kk:16,kk:21} revealed the need for dynamic state verification leading to 
the extension of adding a mutable state. Meanwhile, the embedding of Kripke structures and CTL
into Isabelle have enabled the emulation of Modelchecking and to provide a semantics for attack
trees \cite{kam:17a,kam:17b,kam:17c,kam:18b,kam:19a}.
Attack trees have provided the leverage to integrate Isabelle formal reasoning for IoT systems
as has been illustrated in the CHIST-ERA project SUCCESS \cite{suc:16} where 
attack trees have been used in combination with  the Behaviour Interaction Priority (BIP) component 
architecture model to develop security and privacy enhanced IoT solutions.
This development has emphasized the technical rather than the psychological side of the framework
development and thus branched off the development of the Isabelle {\it Insider} framework into the
Isabelle {\it Infrastructure} framework. Since the strong expressiveness of Isabelle allows to formalize
the IoT scenarios as well as actors and policies, the latter framework can also be applied to 
evaluate IoT scenarios with respect to policies like the European data privacy regulation
GDPR \cite{kam:18a}. Application to security protocols first pioneered in the
auction protocol application \cite{kkp:16,kkp:16a} has further motivated the analysis of Quantum Cryptography
which in turn necessitated the extension by probabilities \cite{kam:19b,kam:19c,kam:19d}.

Requirements raised by these various security and privacy case studies have shown the need for a
cyclic engineering process for developing specifications and refining them towards implementations.
A first case study takes the IoT healthcare application and exemplifies a step-by-step
refinement interspersed with attack analysis using attack trees to increase privacy by ultimately
introducing a blockchain for access control \cite{kam:19a}.
First ideas to support a dedicated security refinement process are available in a preliminary
arxive paper \cite{kam:20a} but the current publication is the first to fully formalize the RR-cycle
and illustrate its application completely on the Corona-virus Warn App (CWA). The earlier workshop publication
\cite{kl:20} provided the formalisation of the CWA illustrating the first two steps but it did not
introduce the fully formalised RR-cycle nor did it apply it to arrive at a solution satisfying the
global privacy policy.

\subsection{Refinement in Formal Methods}
% CSP refinement differences
As has been illustrated on a large scale view %but more informally
in Section \ref{sec:refcomp}, the concept of refinement features in various other formal methods.
The most closely related one is the CSP failure/divergences-refinement as %has been compared
we have seen in the earlier section. Compared to the RR-cycle approach in the Isabelle Infrastructure
framework, CSP offers a range of standardized constructors providing a calculus for event-based specification
refinements. Clearly, CSP has thus a constructive advantage but as has already been discussed in Section
\ref{sec:isacomp}, Isabelle is expressive enough to emulate those constructors. 

In general in CSP, the refinement is focused on events whereas in data-oriented specification formalism,
for example, the B-method \cite{abr:88,abr:96}, %Event-B \cite{abr:10}, (Event-B is more like CSP)
Z and its object-oriented exension Object-Z \cite{sm:00}, refinement means data
refinement, that is, is represented as a relation between abstract and concrete datatypes.
In early works, automated techniques have already been used to derive a concrete specification
from an abstract one provided the concrete data types and refinement relation are given \cite{bl:96,mg:89,ba:88}.
To find a refinement, that is, design the refined specification and define the refinement relation is a
creative and ambigious process; it reflects the development skill of a programmer.
An approach to derive refinement relations and refined specification for Object-Z specifications based on
heuristics is \cite{ks:04}.
By contrast to these earlier work, we make a substantial leap forward as we have a constructive way of
finding a next refinement step guiding this process by attack tree analysis. The resulting RR-cycle thus
provides a systematic way to construct refinements.

% statecharts and there actions; refinement? rather hierarchcial structuring and view of abstraction
% abstraction a la Cousout rather than refinement
Statecharts \cite{har:87} are a formal method focused on graphically modeling system states and their state
transitions and allowing a structured view on the state and its data. Statecharts thus also combine
data in states as well as dynamic behaviour; their semantics resembles our model.
The transitions may be annotated with events that trigger a transition, conditions which guard it, and actions
that are executed when the transition ``fires''. Statecharts may contain data and this data has various layers
of abstraction but this is not a refinement rather a view of the same system at different levels of granularity.
Concerning verification, an embedding of Extended Hierarchical Automata in Isabelle/HOL \cite{hk:01,hk:10} is
related to our work but the focus is on finding property preserving abstractions of the data within the
Statechart \cite{hk:05} to enable model checking. Even though model checking is also involved there, abstraction is
precisely the opposite of refinement. The relationship of Statecharts with CSP and model checking is also investigated
\cite{rw:06} albeit in a dedicated extension of the FDR model checker thus not admitting explicit expression of data
refinement and reasoning on system property preserving specification refinement and general security properties as
we offer here.

\subsection{Dependability and Security Verification}
% Pi-calculus and Proverif 
In software engineering, dependability aggregates the security attributes confidentiality, integrity,
and availability with safety, reliability, and maintainability \cite{alrl:04}. In addition to security, we
explicitly address safety through formal methods and verification with temporal logic since we can explicitly
formalize relevant system properties and prove them on the system specification. However, maintainability
(ability of easy maintenance (repair)) and reliability (continuity of correct service) are provided only
indirectly because specification and formal proof imply higher quality.

Security verification is quite advanced when it comes to the automated analysis of security (or authentication)
protocols. In terms of verification of security using process calculi, the Pi-calculus offers the dedicated model
checker Proverif \cite{Proverif:14}. This work is relevant for our approach since it mechanizes a process calculus
with actor-based communication and provides model checking verification but it does not address the idea of refinement
let alone providing systematic support for dependability engineering. Security protocols have also been formalized
in Isabelle \cite{pau:98} and a similar approach has been shown to integrate well with the Isabelle Insider
framework \cite{kkp:16}.

\section{Conclusions}
\label{sec:concl}
% summary
In this paper we have presented a formalization of the Refinement-Risk cycle (RR-cycle)
within the Isabelle Infrastructure framework which encompasses a notion of formal refinement
with a property preservation theory as well as attack trees based on Kripke structures and CTL.
The RR-cycle iterates the activities of refinement and attack tree analysis until the termination
condition is reached which shows that the global policy is true in the refined model.
The application of this formalization of the RR-cycle is illustrated on the case study
of the Corona-virus Warning App (CWA).
% Future work: Relationship to CSP could inspire dedicated refinement primitives
To contrast the development of the RR-cycle we present an account of related refinement techniques
in other formal methods. The comparison helps to make some of the used concepts clearer for a larger
audience as well as showing up avenues for future work, for example, emulating some of the ready-made
refinement constructors available in process calculi like CSP in the Isabelle Infrastructure framework.

% Advances in guiding the refinement process by attack trees
The real novelty of the RR-cycle approach presented in this paper is guiding the dependability
refinement process by attack trees. 
It is the expressivity of Isabelle's Higher Order logic that permits constructing -- within the logic --
a theory for refinement and attack trees. Thereby, the actual mechanism used in the framework for the
analysis of security applications can be explicitly defined from first principles and its properties
can be mathematically proved in the framework. At the same time, this generic meta-theory can be applied
to application examples. That is, the defined constructs and proved theorems of the Infrastructure
framework can now be instantiated to represent a concrete application example, verify security policies
on it, find attacks using attack trees, and define refinement relations to new models and apply
the RR-cycle constructs to verify -- all this within the same framework and thus with the highest consistency
and correctness guarantees due to both.

\bibliographystyle{IEEEtran} \bibliography{../insider}

% Generated by IEEEtran.bst, version: 1.14 (2015/08/26)
\begin{thebibliography}{10}
\providecommand{\url}[1]{#1}
\csname url@samestyle\endcsname
\providecommand{\newblock}{\relax}
\providecommand{\bibinfo}[2]{#2}
\providecommand{\BIBentrySTDinterwordspacing}{\spaceskip=0pt\relax}
\providecommand{\BIBentryALTinterwordstretchfactor}{4}
\providecommand{\BIBentryALTinterwordspacing}{\spaceskip=\fontdimen2\font plus
\BIBentryALTinterwordstretchfactor\fontdimen3\font minus
  \fontdimen4\font\relax}
\providecommand{\BIBforeignlanguage}[2]{{%
\expandafter\ifx\csname l@#1\endcsname\relax
\typeout{** WARNING: IEEEtran.bst: No hyphenation pattern has been}%
\typeout{** loaded for the language `#1'. Using the pattern for}%
\typeout{** the default language instead.}%
\else
\language=\csname l@#1\endcsname
\fi
#2}}
\providecommand{\BIBdecl}{\relax}
\BIBdecl

\bibitem{alrl:04}
A.~Avizienis, J.~Laprie, B.~Randell, and C.~Landwehr, ``Basic concepts and
  taxonomy of dependable and secure computing,'' \emph{IEEE Transactions on
  Dependable and Secure Computing}, vol.~1, no.~1, pp. 11--33, 2004.

\bibitem{jac:89}
J.~Jacob, ``On the derivation of secure components,'' in \emph{IEEE Security
  and Privacy}.\hskip 1em plus 0.5em minus 0.4em\relax IEEE, 1989, pp.
  242--247.

\bibitem{kl:20}
F.~Kamm\"uller and B.~Lutz, ``Modeling and analyzing the corona-virus warning
  app with the isabelle infrastructure framework,'' in \emph{20th International
  Workshop of Data Privacy Management, DPM’20}, ser. LNCS, vol. 12484.\hskip
  1em plus 0.5em minus 0.4em\relax Springer, 2020, co-located with
  ESORICS’20.

\bibitem{npw:02}
T.~Nipkow, L.~C. Paulson, and M.~Wenzel, \emph{Isabelle/HOL -- A Proof
  Assistant for Higher-Order Logic}, ser. LNCS.\hskip 1em plus 0.5em minus
  0.4em\relax Springer-Verlag, 2002, vol. 2283.

\bibitem{suc:16}
CHIST-ERA, ``Success: Secure accessibility for the internet of things,'' 2016,
  http://www.chistera.eu/projects/success.

\bibitem{kam:19a}
F.~Kamm\"uller, ``Combining secure system design with risk assessment for iot
  healthcare systems,'' in \emph{Workshop on Security, Privacy, and Trust in
  the IoT, SPTIoT’19, colocated with IEEE PerCom}.\hskip 1em plus 0.5em minus
  0.4em\relax IEEE, 2019.

\bibitem{kam:18b}
------, ``Attack trees in isabelle,'' in \emph{20th International Conference on
  Information and Communications Security, ICICS2018}, ser. LNCS, vol.
  11149.\hskip 1em plus 0.5em minus 0.4em\relax Springer, 2018.

\bibitem{hoa:85}
C.~A.~R. Hoare, \emph{Communicating Sequential Processes}.\hskip 1em plus 0.5em
  minus 0.4em\relax Prentice Hall, 1985.

\bibitem{mil:80a}
\BIBentryALTinterwordspacing
R.~Milner, \emph{A Calculus of Communicating Systems}, ser. Lecture Notes in
  Computer Science.\hskip 1em plus 0.5em minus 0.4em\relax Springer, 1980,
  vol.~92. [Online]. Available: \url{https://doi.org/10.1007/3-540-10235-3}
\BIBentrySTDinterwordspacing

\bibitem{mil:99}
------, \emph{Communicating and mobile systems - the Pi-calculus}.\hskip 1em
  plus 0.5em minus 0.4em\relax Cambridge University Press, 1999.

\bibitem{abr:74}
J.~Abrial, ``Data semantics,'' in \emph{Data Base Management, Proceeding of the
  {IFIP} Working Conference Data Base Management, Carg{\`{e}}se, Corsica,
  France, April 1-5, 1974}, J.~W. Klimbie and K.~L. Koffeman, Eds.\hskip 1em
  plus 0.5em minus 0.4em\relax North-Holland, 1974, pp. 1--60.

\bibitem{spy:89}
J.~M. Spivey, \emph{The {Z} notation - a reference manual}, ser. Prentice Hall
  International Series in Computer Science.\hskip 1em plus 0.5em minus
  0.4em\relax Prentice Hall, 1989.

\bibitem{abr:96}
\BIBentryALTinterwordspacing
J.~Abrial, \emph{The B-book - assigning programs to meanings}.\hskip 1em plus
  0.5em minus 0.4em\relax Cambridge University Press, 1996. [Online].
  Available: \url{https://doi.org/10.1017/CBO9780511624162}
\BIBentrySTDinterwordspacing

\bibitem{ros:98}
A.~W. Roscoe, \emph{The Theory and Practice of Concurrency}, ser. Prentice Hall
  Series in Computer Science.\hskip 1em plus 0.5em minus 0.4em\relax Prentice
  Hall, 1998.

\bibitem{cwa:github}
{The Corona-Warn-App Project}, 2020, {https://github.com/corona-warn-app}.

\bibitem{bundes:20}
D.~Bundesregierung, ``{Die Corona-Warn-App: Unterst\"utzt uns im Kampf gegen
  Corona},'' 2020, german government announcement and support of Coronavirus
  warning app, https://www.bundesregierung.de/breg-de/themen/corona-warn-app.

\bibitem{cwa:arch}
{The Corona-Warn-App Project}, ``Corona-warn-app solution architecture,'' 2020,
  {https://github.com/corona-warn-app/cwa-documentation/blob/master/solution\_architecture.md}.

\bibitem{dp3t:github}
{The DP-3T Project}, ``{Decentralized Privacy-Preserving Proximity Tracing},''
  2020, https://github.com/DP-3T.

\bibitem{enf:proj}
{Apple and Google}, ``Exposure notification framework,'' 2020,
  {https://www.google.com/covid19/exposurenotifications/}.

\bibitem{pepppt:github}
{The PEPP-PT Project}, ``{Pan-European Privacy-Preserving Proximity Tracing},''
  2020, https://github.com/PEPP-PT.

\bibitem{pepppt:ROBERT}
{The ROBERT Project}, ``{ROBust and privacy-presERving proximity Tracing
  protocol},'' 2020, https://github.com/ROBERT-proximity-tracing.

\bibitem{pepppt:dp3tana}
S.~Vaudenay, ``{Analysis of DP3T: Between Scylla and Charybdis},'' 2020,
  {https://eprint.iacr.org/2020/399.pdf}.

\bibitem{pepppt:dp3tresp}
{The DP-3T Project}, ``Response to {'Analysis of DP3T: Between Scylla and
  Charybdis'},'' 2020, {https://github.com/DP-3T/documents/blob/master/Security
  analysis/Response to 'Analysis of DP3T'.pdf}.

\bibitem{dp3t:psre}
------, ``Privacy and security risk evaluation of digital proximity tracing
  systems,'' 2020, {https://github.com/DP-3T/documents/blob/master/Security
  analysis/Privacy and Security Attacks on Digital Proximity Tracing
  Systems.pdf}.

\bibitem{kam:20gitsc}
\BIBentryALTinterwordspacing
F.~Kamm\"uller, ``Isabelle infrastructure framework for ibc,'' 2020, isabelle
  sources for IBC formalisation. [Online]. Available:
  \url{https://github.com/flokam/IsabelleSC}
\BIBentrySTDinterwordspacing

\bibitem{dp3t:wp}
{The DP-3T Project}, ``Decentralized privacy-preserving proximity tracing -
  {White Paper},'' 2020, https://github.com/DP-3T/documents/blob/master/DP3T
  White Paper.pdf.

\bibitem{kp:13}
F.~Kamm\"uller and C.~W. Probst, ``Invalidating policies using structural
  information,'' in \emph{IEEE Security and Privacy Workshops, Workshop on
  Research in Insider Threats, WRIT'13}, 2013.

\bibitem{kp:14}
------, ``Combining generated data models with formal invalidation for insider
  threat analysis,'' in \emph{IEEE Security and Privacy Workshops, Workshop on
  Research in Insider Threats, WRIT'14}, 2014.

\bibitem{kp:16}
\BIBentryALTinterwordspacing
------, ``Modeling and verification of insider threats using logical
  analysis,'' \emph{IEEE Systems Journal, Special issue on Insider Threats to
  Information Security, Digital Espionage, and Counter Intelligence}, vol.~11,
  no.~2, pp. 534--545, 2017. [Online]. Available:
  \url{http://dx.doi.org/10.1109/JSYST.2015.2453215}
\BIBentrySTDinterwordspacing

\bibitem{cmt:12}
\BIBentryALTinterwordspacing
D.~M. Cappelli, A.~P. Moore, and R.~F. Trzeciak, \emph{{The CERT Guide to
  Insider Threats: How to Prevent, Detect, and Respond to Information
  Technology Crimes (Theft, Sabotage, Fraud)}}, 1st~ed., ser. SEI Series in
  Software Engineering.\hskip 1em plus 0.5em minus 0.4em\relax Addison-Wesley
  Professional, Feb. 2012. [Online]. Available:
  \url{http://www.amazon.com/exec/obidos/redirect?tag=citeulike07-20\&path=ASIN/0321812573}
\BIBentrySTDinterwordspacing

\bibitem{kkp:16}
F.~Kamm\"uller, M.~Kerber, and C.~Probst, ``Towards formal analysis of insider
  threats for auctions,'' in \emph{8th ACM CCS International Workshop on
  Managing Insider Security Threats, MIST’16}.\hskip 1em plus 0.5em minus
  0.4em\relax ACM, 2016.

\bibitem{kkp:16a}
\BIBentryALTinterwordspacing
------, ``Insider threats for auctions: Formal modeling, proof, and certified
  code,'' \emph{Journal of Wireless Mobile Networks, Ubiquitous Computing, and
  Dependable Applications (JoWUA)}, vol.~8, no.~1, 2017. [Online]. Available:
  \url{http://doi.org/10.22667/JOWUA.2017.03.31.044}
\BIBentrySTDinterwordspacing

\bibitem{pau:98}
L.~C. Paulson, ``The inductive approach to verifying cryptographic protocols,''
  \emph{Journal of Computer Security}, vol.~6, no. 1-2, pp. 85--128, 1998.

\bibitem{kk:16}
F.~Kamm\"uller and M.~Kerber, ``Investigating airplane safety and security
  against insider threats using logical modeling,'' in \emph{IEEE Security and
  Privacy Workshops, Workshop on Research in Insider Threats, WRIT'16}.\hskip
  1em plus 0.5em minus 0.4em\relax IEEE, 2016.

\bibitem{kk:21}
\BIBentryALTinterwordspacing
------, ``Applying the isabelle insider framework to airplane security,''
  \emph{Science of Computer Programming}, vol. 206, 2021. [Online]. Available:
  \url{https://doi.org/10.1016/j.scico.2021.102623}
\BIBentrySTDinterwordspacing

\bibitem{kam:17a}
F.~Kamm\"uller, ``A proof calculus for attack trees,'' in \emph{Data Privacy
  Management, DPM’17, 12th Int. Workshop}, ser. LNCS, vol. 10436.\hskip 1em
  plus 0.5em minus 0.4em\relax Springer, 2017, co-located with ESORICS’17.

\bibitem{kam:17b}
------, ``Human centric security and privacy for the iot using formal
  techniques,'' in \emph{3d International Conference on Human Factors in
  Cybersecurity}, ser. Advances in Intelligent Systems and Computing, vol.
  593.\hskip 1em plus 0.5em minus 0.4em\relax Springer, 2017, pp. 106--116,
  affiliated with AHFE’2017.

\bibitem{kam:17c}
------, ``Formal models of human factors for security and privacy,'' in
  \emph{5th International Conference on Human Aspects of Security, Privacy and
  Trust, HCII-HAS 2017}, ser. LNCS, vol. 10292.\hskip 1em plus 0.5em minus
  0.4em\relax Springer, 2017, pp. 339--352, affiliated with HCII 2017.

\bibitem{kam:18a}
------, ``Formal modeling and analysis of data protection for gdpr compliance
  of iot healthcare systems,'' in \emph{IEEE Systems, Man and Cybernetics,
  SMC2018}.\hskip 1em plus 0.5em minus 0.4em\relax IEEE, 2018.

\bibitem{kam:19b}
\BIBentryALTinterwordspacing
------, ``Qkd in isabelle – bayesian calculation,'' \emph{arXiv}, vol. cs.CR,
  2019. [Online]. Available: \url{https://arxiv.org/abs/1905.00325}
\BIBentrySTDinterwordspacing

\bibitem{kam:19c}
\BIBentryALTinterwordspacing
------, ``Attack trees in isabelle extended with probabilities for quantum
  cryptography,'' \emph{Computer \& Security}, vol.~87, 2019. [Online].
  Available: \url{//doi.org/10.1016/j.cose.2019.101572}
\BIBentrySTDinterwordspacing

\bibitem{kam:19d}
\BIBentryALTinterwordspacing
------, ``Formalizing probabilistic quantum security protocols in the isabelle
  infrastructure framework,'' informal Presentation at Computability in Europe,
  CiE 2019. [Online]. Available:
  \url{https://www.aemea.org/CIE2019/CIE_2019_Abstracts.pdf#page=35}
\BIBentrySTDinterwordspacing

\bibitem{kam:20a}
F.~Kammüller, ``A formal development cycle for security engineering in
  isabelle,'' 2020.

\bibitem{abr:88}
\BIBentryALTinterwordspacing
J.~Abrial, ``The {B} tool (abstract),'' in \emph{{VDM} '88, {VDM} - The Way
  Ahead, 2nd VDM-Europe Symposium, Dublin, Ireland, September 11-16, 1988,
  Proceedings}, ser. Lecture Notes in Computer Science, R.~E. Bloomfield, L.~S.
  Marshall, and R.~B. Jones, Eds., vol. 328.\hskip 1em plus 0.5em minus
  0.4em\relax Springer, 1988, pp. 86--87. [Online]. Available:
  \url{https://doi.org/10.1007/3-540-50214-9\_8}
\BIBentrySTDinterwordspacing

\bibitem{sm:00}
G.~Smith, \emph{The Object-Z Specification Language}.\hskip 1em plus 0.5em
  minus 0.4em\relax Kluwer Academic Publishers, 2000.

\bibitem{bl:96}
M.~J. Butler and T.~L{\aa}ngbacka, ``Program derivation using the refinement
  calculator,'' in \emph{Theorem Proving in Higher Order Logics, TPHOLs'96},
  ser. LNCS, J.~von Wright, J.~Grundy, and J.~Harrison, Eds., vol. 1125.\hskip
  1em plus 0.5em minus 0.4em\relax Springer, 1996, pp. 93--108.

\bibitem{mg:89}
C.~C. Morgan and P.~H.~B. Gardiner, ``Data refinement by calculation,''
  \emph{Acta Informatica}, vol.~27, no.~6, pp. 481--503, 1989.

\bibitem{ba:88}
R.~J.~R. Back, ``A calculus of refinements for program derivation,'' \emph{Acta
  Informatica}, vol.~25, no.~6, pp. 593--624, 1988.

\bibitem{ks:04}
F.~Kamm\"uller and J.~W. Sanders, ``Heuristics for refinement relations,'' in
  \emph{Second Interational Conference of Software Engineering and Formal
  Methods, SEFM'04}.\hskip 1em plus 0.5em minus 0.4em\relax IEEE, 2004.

\bibitem{har:87}
D.~Harel, ``Statecharts: A visual formalism for complex systems,''
  \emph{Science of Computer Programming}, vol.~8, pp. 231--274, 1987.

\bibitem{hk:01}
\BIBentryALTinterwordspacing
S.~Helke and F.~Kamm{\"{u}}ller, ``Representing hierarchical automata in
  interactive theorem provers,'' in \emph{Theorem Proving in Higher Order
  Logics, 14th International Conference, TPHOLs 2001, Edinburgh, Scotland, UK,
  September 3-6, 2001, Proceedings}, ser. Lecture Notes in Computer Science,
  R.~J. Boulton and P.~B. Jackson, Eds., vol. 2152.\hskip 1em plus 0.5em minus
  0.4em\relax Springer, 2001, pp. 233--248. [Online]. Available:
  \url{https://doi.org/10.1007/3-540-44755-5\_17}
\BIBentrySTDinterwordspacing

\bibitem{hk:10}
\BIBentryALTinterwordspacing
------, ``Formalizing statecharts using hierarchical automata,'' \emph{Arch.
  Formal Proofs}, vol. 2010, 2010. [Online]. Available:
  \url{https://www.isa-afp.org/entries/Statecharts.shtml}
\BIBentrySTDinterwordspacing

\bibitem{hk:05}
\BIBentryALTinterwordspacing
------, ``Structure preserving data abstractions for statecharts,'' in
  \emph{Formal Techniques for Networked and Distributed Systems - {FORTE} 2005,
  25th {IFIP} {WG} 6.1 International Conference, Taipei, Taiwan, October 2-5,
  2005, Proceedings}, ser. Lecture Notes in Computer Science, F.~Wang, Ed.,
  vol. 3731.\hskip 1em plus 0.5em minus 0.4em\relax Springer, 2005, pp.
  305--319. [Online]. Available: \url{https://doi.org/10.1007/11562436\_23}
\BIBentrySTDinterwordspacing

\bibitem{rw:06}
B.~Roscoe and Z.~Wu, ``Verifying statemate statecharts using {CSP} and {FDR},''
  in \emph{Proceedings of International Conference on Formal Engineering
  Methods (ICFEM)}, ser. LNCS, Z.~Liu and J.~He, Eds., vol. 4260.\hskip 1em
  plus 0.5em minus 0.4em\relax Springer-Verlag, 2006.

\bibitem{Proverif:14}
B.~Blanchet, ``Automatic verification of security protocols in the symbolic
  model: the verifier {P}ro{V}erif,'' in \emph{Foundations of Security Analysis
  and Design VII, FOSAD Tutorial Lectures}, ser. Lecture Notes on Computer
  Science, A.~Aldini, J.~Lopez, and F.~Martinelli, Eds.\hskip 1em plus 0.5em
  minus 0.4em\relax Springer Verlag, 2014, vol. 8604, pp. 54--87.

\end{thebibliography}

%\appendix

%\vspace{.3cm}
%\centerline{\sc X. Author's Biography}
%\vspace{.2cm}
%
%Florian Kamm\"uller is an Associate Professor in the Department of 
%Computer Science at Middlesex University London and Privatdozent (honorary professor)
%at Technische Universitaet Berlin. He holds a PhD from the University of Cambridge 
%on "Modular Reasoning in Isabelle" and a Habilitation from Technische 
%Universitaet Berlin on "Interactive Theorem Proving in Software Engineering". 
%His research is centered around applying interactive theorem proving
%and other formal methods to applications from software and security engineering.
%He has led and contributed to several research projects in Germany and the UK
%has been leading the recent European CHIST-ERA project SUCCESS on security and 
%privacy for the IoT where formal security engineering is applied to support 
%cost-efficient transparent health care monitoring systems with follow up
%project in the domain of sustainable network slicing currently being launched.

\end{document}